\documentclass[prb,twocolumn,amsmath,amssymb,superscriptaddress,floatfix]{revtex4}

\pdfoutput=1

\usepackage{amsmath}
\usepackage{amsfonts}
\usepackage{float}
\usepackage{ graphicx } 
\usepackage{ bm } 
\usepackage{ color }

\usepackage[dvipsnames]{xcolor}
\usepackage{mathrsfs}

\newcommand{\be}{\begin{equation}}
\newcommand{\ee}{\end{equation}}
\newcommand{\bea}{\begin{eqnarray}}
\newcommand{\eea}{\end{eqnarray}}

\begin{document}

\title{Topological states in chiral electronic chains}

\author{P. D. Sacramento}
\author{M. F. Madeira}
\affiliation{Departamento de F\'{\i}sica and CeFEMA, Instituto Superior T\'{e}cnico, Universidade de Lisboa, Av. Rovisco Pais, 1049-001 Lisboa, Portugal}
\date{\today}

\begin{abstract}
We consider the influence of topological phases, or their vicinity, on the spin density and spin
polarization through a chiral chain.
We show the quantization of the Berry phase
in a one-dimensional polarization helix structure, under the presence of an external
magnetic field, and show its influence on the
spin density. The polar angle of the momentum space spin density 
becomes quantized in the regime that the Berry phase is quantized, as a result of the 
combined effect of the induced spin-orbit coupling and the external transverse magnetic
field, while
the edge states do not show the polar angle quantization,
in contrast with the bulk states.
Under appropriate conditions, the model can be generalized to
have similarities with a chain with non-homogeneous Rashba spin-orbit couplings, 
with zero or low energy edge states. 
Due to the breaking of time-reversal symmetry, we recover the effect of chiral induced spin polarization
and spin transport across the chiral chain, when coupling to external leads. Some consequences
of the quantized spin polarization and low energy states 
on the spin transport are discussed. 
\end{abstract}


\maketitle

\section{Introduction}

Chiral structures have attracted considerable interest. Examples are chiral magnets, chiral metals or  
chiral molecules, for
instance in the context of chiral induced selective spin transport \cite{bauer}. Molecules
such as DNA or similar structures, either single stranded or doubly
stranded, have been studied 
\cite{xie,guo1,gutierrez1,eremko,gutierrez2,gutierrez3,naaman,varela,caetano,matityahu,utsumi,inui,evers}.
Even though some controversy is still unresolved on a proper description of the
large effect found experimentally for the spin polarization through specific chiral structures
\cite{evers}, there is a variety of proposals, from an effect due to the spin of the 
electrons \cite{xie,guo1,gutierrez1,eremko,gutierrez2,gutierrez3,naaman,varela,caetano},
due to an orbital contribution \cite{liu}, or due to a mixture of both \cite{utsumi}. 
The requirement of time-reversal symmetry
breaking is implemented in different ways, either through an explicit model without this
symmetry (as effectively initial models considered) or through the effect of an external magnetic field,
via non-unitary effects \cite{matityahu}, coupling to orbital degrees of freedom \cite{utsumi} 
or taking into account effects of
interactions \cite{fransson}. Synthetic structures with intrinsic or applied chiral
electric fields may also be of interest.

The chiral structure may arise due to a twisting polarization (and associated
twisting electric field) along the molecule, such as in the DNA-type molecule or in bent-core
molecules \cite{lorman1,selinger,lorman2,mettout1} 
in the context of liquid crystals. The interest is in the electronic response to
the presence of this chiral structure and it may be relevant when the electronic density
and mobility are significant. While in equilibrium an effect is not expected, the chiral
molecule affects the spin density of a moving electron when attaching
leads at the extremes of the structure.

One considers therefore an electron moving in the presence of a twisting electric field 
that couples to the spin and leads to a non-homogeneous spin-orbit like effect.
Typically, the models considered have a gapless structure that may be changed by adding
gaps to the system, either adding a magnetic field (which naturally breaks time
reversal invariance and may lead to spin transport) or introducing some form of distortion
of the lattice (like in the SSH model \cite{SSH}), adding local potential energy terms that separate the energies of the
various lattice sites inside the unit cell of the helix-like structure. 

A question that may arise is whether the considered systems may exhibit a topological nature. 
The change from a gapless to a gapped regime will possibly facilitate the emergence
and analysis of possible topological non-trivial states.
In a gapless regime, edge states are usually not found, even though
exceptions have been identified in gapless critical systems \cite{Verresen,prr3}.
In a gapped regime edge states may appear, and one may wonder if signals 
of their presence will be felt on spin transport.
One may also ask how the spin density and spin polarization are affected.
The presence of the terms that lead to a gapped system, such as the addition of an
external magnetic field or the lattice distortion, may change the symmetry of 
the problem, and the possible edge states 
may be zero energy modes or finite energy modes \cite{ssh1,ssh2}.
The appearance of topology in the context of chiral molecules has been proposed before
\cite{liu}, including as a result of a time-dependent perturbation \cite{guo2,schuster}.

\begin{figure}
\centering
\includegraphics[width=0.26\textwidth]{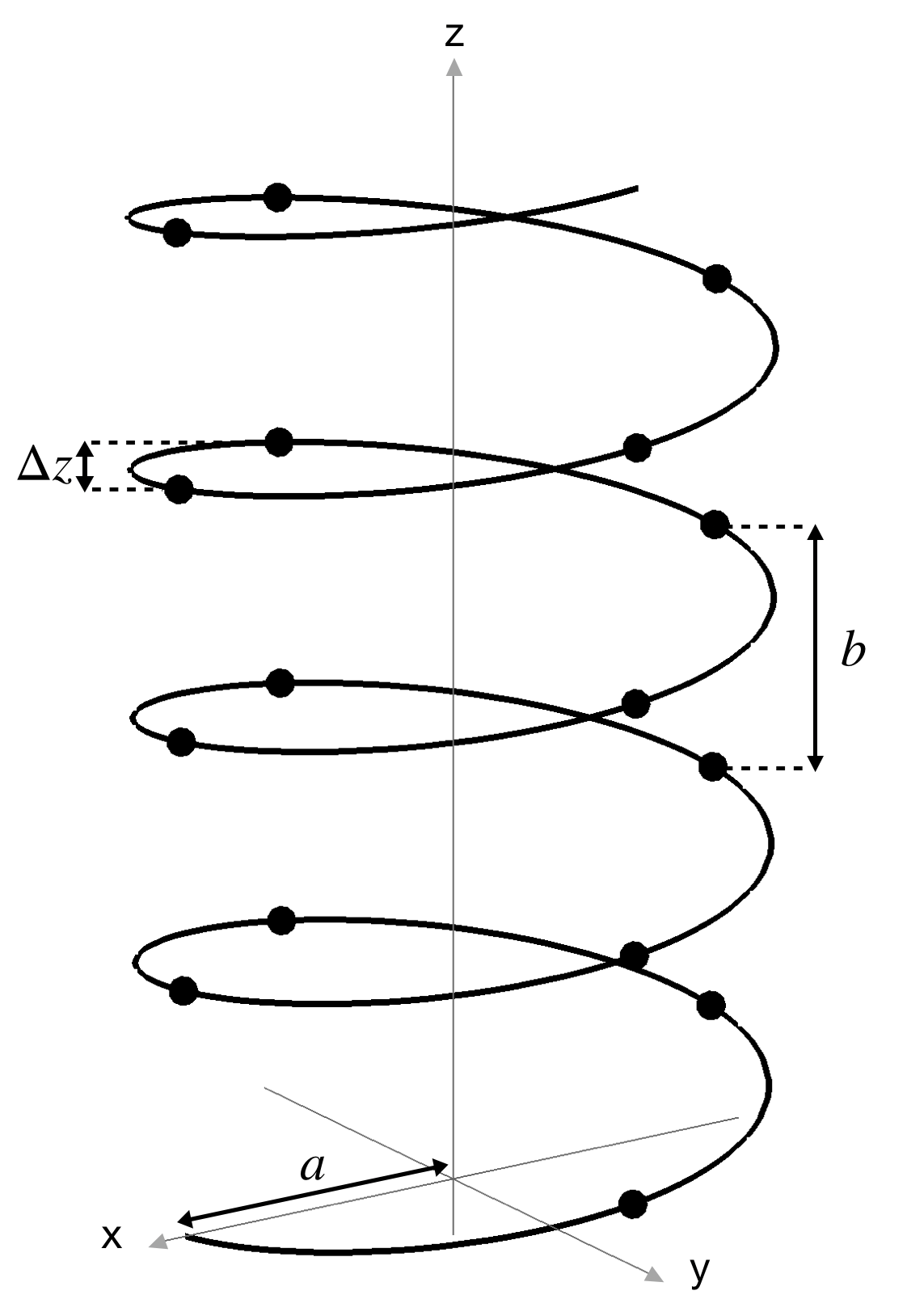}
\includegraphics[width=0.16\textwidth]{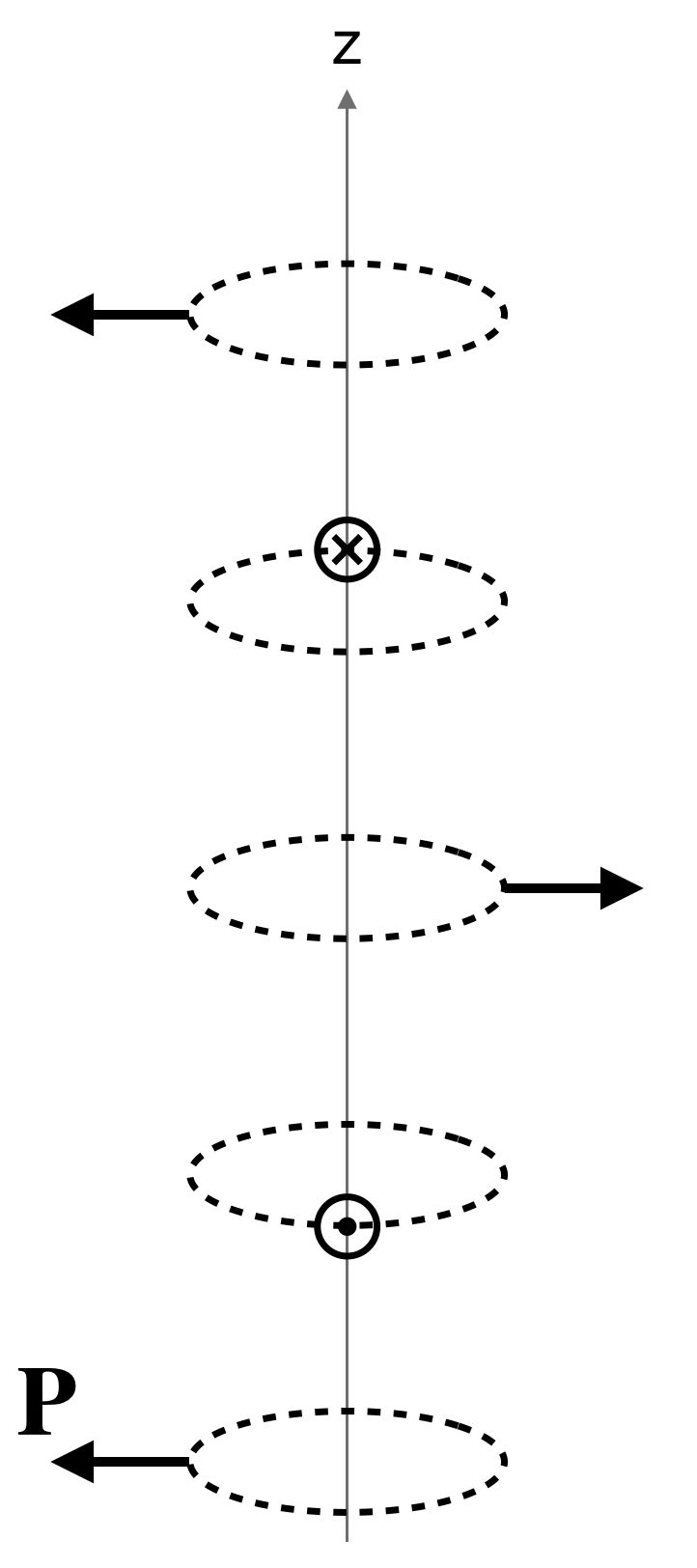}
\caption{\label{bent}
Schematic drawing of the three-dimensional helix structure
with atoms or molecules at locations
$\mathbf{r}(m,\alpha) = \left( a \cos \frac{2 \pi}{\gamma} \alpha, a \sin \frac{2 \pi}{\gamma} \alpha, 
\left[ (m-1) \gamma + \alpha \right] \Delta z \right)$,
along an helix of radius $a$, where $\alpha=1,\cdots ,\gamma$, $b$ is the helix pitch, $\gamma$ is the number
of molecules in each pitch and $m$ is an index of a given set of molecules in a given pitch. 
Finally, $\Delta z=b/\gamma$ is the spacing along the direction of the helix, $z$, between consecutive molecules.
On the right we show a scheme of the polarization vectors taken in the plane perpendicular to the
molecule axis, taking as example $\gamma=4$.
}
\end{figure}

In this work we consider electronic states that
propagate along a chiral chain, in a way similar to previous treatments. 
An example is a helix composed of molecules or atoms that display
polarization and an associated electric field, $\mathbf{E}$. As an electron propagates
along the helix, the electric field twists in a helical arrangement. 
This is, in general, a three-dimensional process. 
The polarization profile gives rise to an electric field that also has an helix-like structure, as shown
in Fig. \ref{bent}. The helix may be divided in unit cells with $\gamma$ atoms that are displaced along
the axis of the helix, $z$-axis, by an amount $\Delta z=b/\gamma$, where $b$ is the helix pitch. 
The radius of the helix is $a$. In Fig. \ref{bent} the helix is represented
for the case of $\gamma=4$.

Here, to simplify, we consider one-dimensional rods, that are immersed in the three-dimensional structure.
At each molecule or atom location in a given unit cell, $m$, there is a polarization given by
\be
\mathbf{P}(\alpha) = \left( P \cos \frac{2 \pi}{\gamma} \alpha, P \sin \frac{2 \pi}{\gamma} \alpha, 
0 \right)
\ee
where $\alpha=1,\cdots, \gamma$ labels each atom within the unit cell.
The problem is simplified considering an electron that moves in the direction of the
axis of the helix and, therefore, we restrict its momentum to the $z$ direction.  
In this simplification the momentum of the electron is $\mathbf{k}=k \mathbf{e}_z$. Due to the
presence of the electric field, the electron feels in its reference frame an effective magnetic
field in the $x-y$ plane, that couples to the electron spin through a Zeeman-like effect. There is 
therefore a spin-orbit-like contribution to the Hamiltonian of the electron of the form
$H_{SO} = \lambda \boldsymbol{\sigma} \cdot \left( \mathbf{k} \times \mathbf{E} \right)$,
where $\lambda$ parametrizes the amplitude of the spin-orbit-like coupling.
The chiral nature of the electric field configuration is therefore encoded in
this spin-orbit term.
A term like this has been shown to give rise to spin selective transport across a
chiral molecule, if time reversal symmetry is broken. 
In this work we add an external magnetic field.
We show that electronic bands arise with a gapped structure and with non-trivial quantized Berry (Zak) phases
\cite{zak},
under some conditions. We calculate the spin density and show some quantization of the spin density direction
when the Berry phase is quantized, either taking the average of the spin density on momentum space
eigenstates or taking the average over real space bulk states . In general, 
there are edge states of finite energy that, however, do not
show spin density quantization. 

We consider some generalizations of the model that give origin
to additional topological zero energy edge states, in the absence of magnetic field.
These are however shifted to finite, small, energies when the magnetic field is
applied. It turns out that the quantization of the Zak phase is lost in some of the generalized models.
The model without magnetic field or lattice distortion is topological and in class $CII$, but is gapless.
Adding the lattice distortion the model is still topological and becomes gapped.
Adding the magnetic field the model symmetry changes to class $C$, which is not topological in $1d$.
Combining the effects of the lattice distortion and the magnetic field, the model has no time-reversal
symmetry and displays finite energy edge states. The various cases will be detailed ahead. 
Coupling the chiral structure to external leads, 
a spin polarization is found considering the 
cases when there is time-reversal symmetry breaking. Some enhancement of the spin polarization
is found at low energies due to the egde states and in the regime where the Berry phase is
quantized, associated with the preferred spin density direction. 

In Section II we consider a tight-binding model with spin-orbit coupling and external magnetic field
and study its energy spectrum and symmetries, calculate the Berry phase of the energy bands
and determine the conditions for its quantization. In Section III we calculate the momentum space
spin density and associate a quantization property with the quantization of the Berry phase.
In Section IV a real-space description is adopted to determine the edge states and the real
space spin density. In Section V we consider some generalizations of the model that lead
to zero energy or low energy edge modes. In Section VI we study the spin transport
along the system, coupling the chain to leads.

\section{Energy bands and Berry phase in a one-dimensional chiral structure}


We consider a tight-binding model for the motion of the electrons. The spin-orbit coupling
leads to a term that has to be symmetrized appropriately \cite{gutierrez1,gutierrez2,pareek}.
The Hamiltonian in real space is chosen to be
\bea
H &=& \sum_j \sum_{m_s} c_{j,m_s}^{\dagger} \left( -\mu + \mu(j)\right) c_{j,m_s}
\nonumber \\
 &-& \sum_j \sum_{m_s,m_s^{\prime}} c_{j,m_s}^{\dagger} \left( \mathbf{B} \cdot \boldsymbol{\sigma}_{m_s,m_s^{\prime}}
\right) c_{j,m_s^{\prime}}
\nonumber \\
&-& w \sum_j \sum_{m_s} \left( c_{j,m_s}^{\dagger} c_{j+1,m_s} + c_{j+1,m_s}^{\dagger} c_{j,m_s} \right)
\nonumber \\
&-& \lambda \sum_j 
\left( c_{j+1,\uparrow}^{\dagger}, c_{j+1,\downarrow}^{\dagger} \right)
\left(\begin{array}{cc}
0 & \chi_j \\
-\chi_j^* & 0
\end{array}\right)
\left(\begin{array}{c}
c_{j,\uparrow} \\
c_{j,\downarrow}
\end{array}\right)
\nonumber \\
&-& \lambda \sum_j 
\left( c_{j,\uparrow}^{\dagger}, c_{j,\downarrow}^{\dagger} \right)
\left(\begin{array}{cc}
0 & -\chi_j \\
\chi_j^* & 0
\end{array}\right)
\left(\begin{array}{c}
c_{j+1,\uparrow} \\
c_{j+1,\downarrow}
\end{array}\right)
\label{Hamiltonian}
\eea
where $c_{j,m_s} = c_{m,\alpha,m_s}$ destroys an electron at site $j$ and with spin projection
$m_s$. 
Here $j=(m-1)\gamma+\alpha$ labels each molecule or atom location, in cell $m$
and location $\alpha$ within the cell, $\mu$ is the chemical potential,
$w$ is the hopping amplitude to a nearest neighbor location,
$m_s=\uparrow, \downarrow$ are the spin projections
and $\chi(j)=i E_y(j)-E_x(j)= -E e^{-i \Phi (j-1)}$ has a constant amplitude and where $\Phi=(2\pi)/\gamma$. 
($\mu,w,B,\lambda E$ have energy units).
The amplitude of the electric field is taken as $E=\sqrt{E_x^2+E_y^2}=1$.
Also, we have added a local external magnetic
field, $\mathbf{B}$, that couples to the spin density of the electrons.
This term explicitly breaks time reversal invariance and allows the opening of gaps between
the bands. Also, it allows the study of the effect of time-reversal symmetry breaking on a multiband
fermionic system \cite{ssh2}.

If the chiral chain is finite the electric field amplitude is not a constant.
In order to simplify the problem we consider a long chain and neglect the effect of
the ends of the chain by considering a constant amplitude throughout. 
We take $w=1, \mu=0$. 
Also, in a more
realistic model there is coupling between sites $j$ and $j+\gamma$, which we neglect
here, as well as the three-dimensional motion of the electrons along the structure.
In addition, we consider a simplified model of one orbital state in which the hopping term is a constant 
and consider that the effect of the twist along the helix only affects the spin-orbit
term \cite{gutierrez1,gutierrez3}.

The effects of a spin-orbit interaction in multiband systems have been extensively studied
in the context of topological insulators \cite{hasan,zhang,kane} focusing mainly on homogeneous
spin-orbit terms. Non-homogeneous (or modulated) spin-orbit terms have also been considered
such as due to a curved wire \cite{gentile} or in the generalized Aubry-Andr\'e class \cite{malard}.
Here we focus on a spin-orbit interaction that encodes the chirality of the polarization or electric
field configuration.

The explicit construction of the states is obtained diagonalizing the Hamiltonian in
real space for a finite helix using open boundary conditions, giving both the bulk states and edge states.
Considering an infinite helix we may use a momentum space description,
with a basis of size $\gamma$. 
The Hamiltonian in momentum space is a $(\gamma \times 2) \times (\gamma \times 2)$ matrix that gives rise
to $\gamma \times 2$ electronic bands. 
In the case of $\gamma=2$ 
the electric field has a staggered structure and so we consider higher values of $\gamma$. 

Let us consider explicitly, as an example, the case of $\gamma=4$. 
In this case the Hamiltonian matrix in momentum space is written
in a basis of vectors $(1 \uparrow, 1 \downarrow, 3 \uparrow, 3 \downarrow, 
2 \uparrow, 2 \downarrow, 4 \uparrow, 4 \downarrow)^T$.
Considering that in this case
\be
\chi=(-E,i E, E, -i E)
\ee
and defining $Z_{ijk}=\sigma_i \otimes
\sigma_j \otimes \sigma_k$, we can write the Hamiltonian matrix as
\bea
H(k)  & = &  -w Z_{100} -\frac{w}{2} \left( 1+\cos k \right) Z_{110}
\nonumber \\
&-& \frac{w}{2} \left( 1 - \cos k \right) Z_{220}
-\frac{w}{2} \sin k \left( Z_{120} + Z_{210} \right)
\nonumber \\
&+& \lambda E \left( Z_{232} -\frac{1}{2} Z_{121}+\frac{1}{2} Z_{211} \right)
\nonumber \\
&+& \frac{\lambda E}{2} \left( \sin k \left( Z_{111}-Z_{221} \right)
-\cos k \left( Z_{121}+Z_{211} \right) \right)
\nonumber \\
&-& \left( B_x Z_{001} +B_y Z_{002} + B_z Z_{003} \right)
\eea

\subsection{Symmetries}

The Hamiltonian matrix has some symmetries. Consider for instance the case of $\gamma=4$.
In this case the Hamiltonian has particle-hole (or charge conjugation) symmetry, defined by an operator, $C$,
that satisfies $C H^*(-k) C^{\dagger}=-H(k)$, 
which yields $C= \left(\sigma_3 \otimes \sigma_0 \otimes \sigma_2 \right)$.
Since $C^T=-C$ the model is in class C, which in one dimension is predicted to be non-topological
\cite{schnyder}.
If the external magnetic field is absent, $\mathbf{B}=0$, then the system also has time reversal
symmetry, $T H^*(-k) T^{\dagger}=H(k)$, with $T=\left( \sigma_0 \otimes \sigma_0 \otimes \sigma_2 \right)$ and
chiral symmetry, $O H(k) O^{\dagger} =-H(k)$, with $O = \left( \sigma_3 \otimes \sigma_0 \otimes \sigma_0 \right)$.
In this case the model is in class CII and in one dimension is predicted to be a class $Z$ topological system.
Note that the model is gapless in this case.

In addition to the global symmetries considered, the Hamiltonian in real space also has a spatial symmetry, in some
conditions. Defining a transformation of the fermion operators as
\be
c_{j,m_s} \rightarrow c_{j+1,m_s} e^{i m_s \frac{\pi}{\gamma}}
\ee
the spin-orbit term and the hopping term remain invariant. Including an external magnetic
field along the $z$ direction also leaves the Hamiltonian invariant (under periodic boundary conditions).
However, the $B_x$ and $B_y$ components do not remain invariant. The term due to $B_x$ transforms as
\be
c_{j,\uparrow}^{\dagger} B_x c_{j,\downarrow} + c_{j,\downarrow}^{\dagger} B_x c_{j,\uparrow} 
\rightarrow
c_{j,\uparrow}^{\dagger} B_x c_{j,\downarrow} e^{-i \frac{2\pi}{\gamma}} 
+ c_{j,\downarrow}^{\dagger} B_x c_{j,\uparrow} e^{i \frac{2\pi}{\gamma}}
\ee
and the term due to $B_y$ transforms as
\bea
& & c_{j,\uparrow}^{\dagger} (-i B_y) c_{j,\downarrow} + c_{j,\downarrow}^{\dagger} (i B_y) c_{j,\uparrow} 
\nonumber \\
\rightarrow
& & c_{j,\uparrow}^{\dagger} (-i B_y) c_{j,\downarrow} e^{-i \frac{2\pi}{\gamma}} 
+ c_{j,\downarrow}^{\dagger} (i B_y) c_{j,\uparrow} e^{i \frac{2\pi}{\gamma}}
\eea

\begin{figure}
\centering
\includegraphics[width=0.34\textwidth]{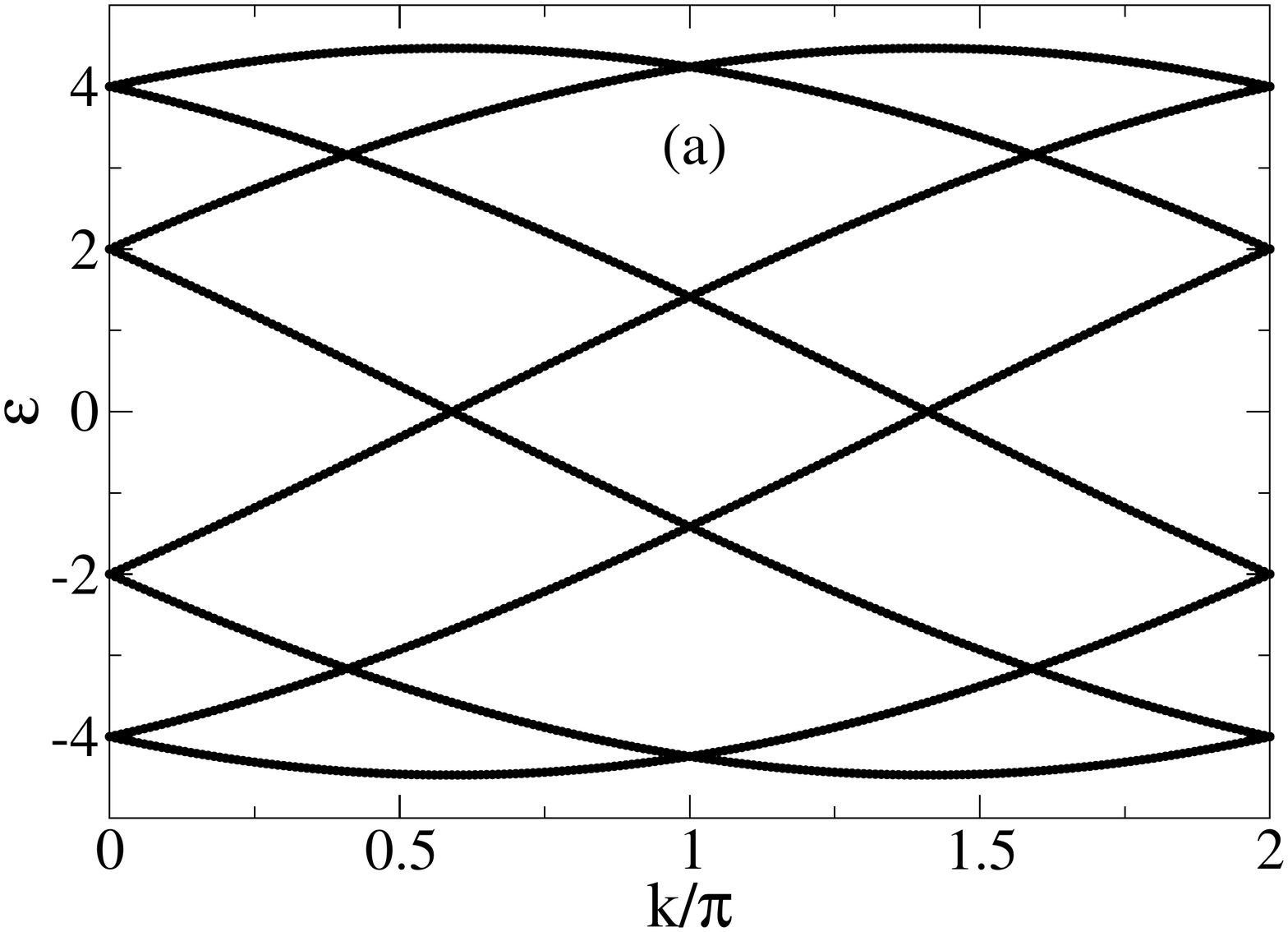}
\includegraphics[width=0.34\textwidth]{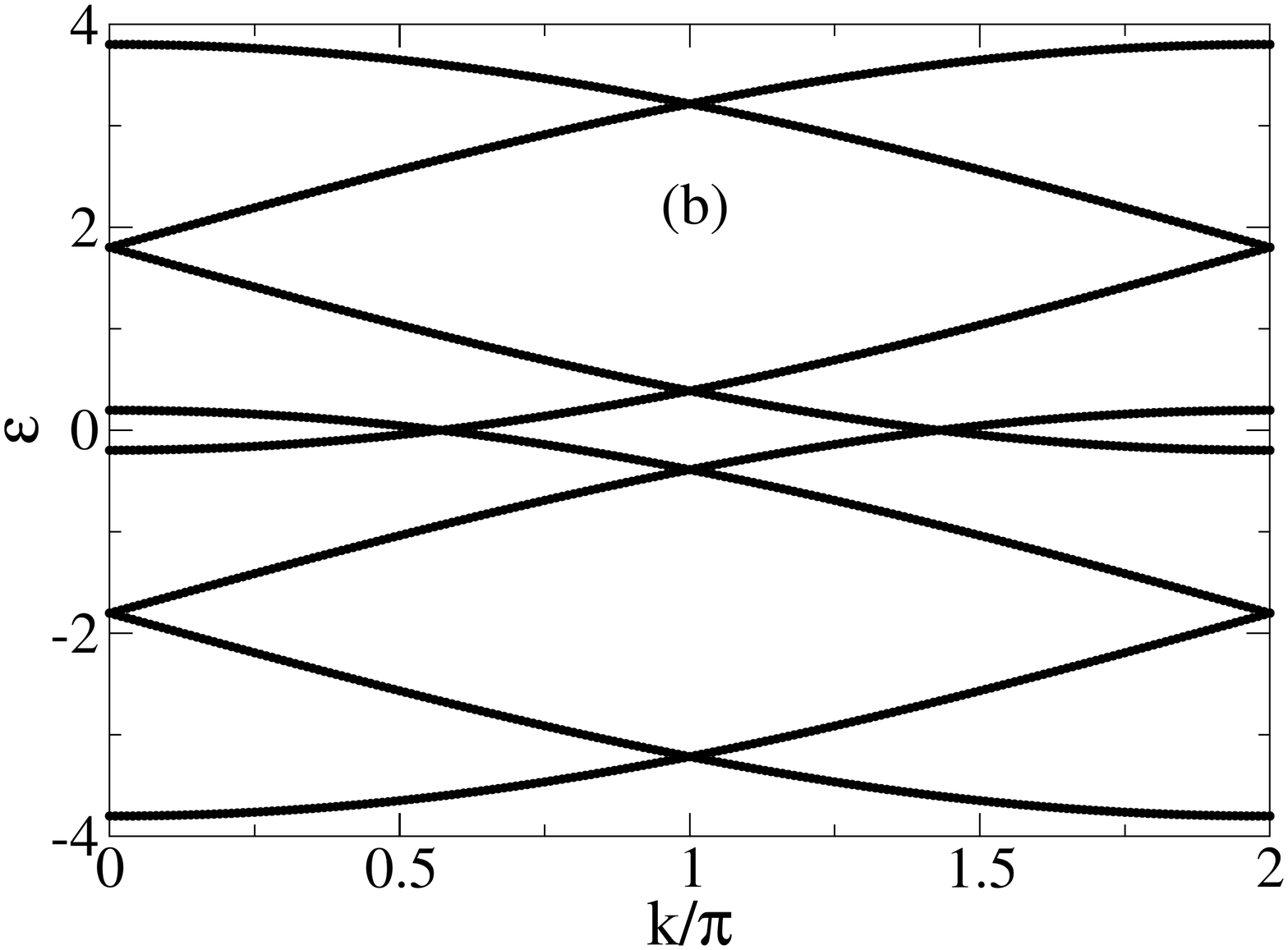}
\includegraphics[width=0.34\textwidth]{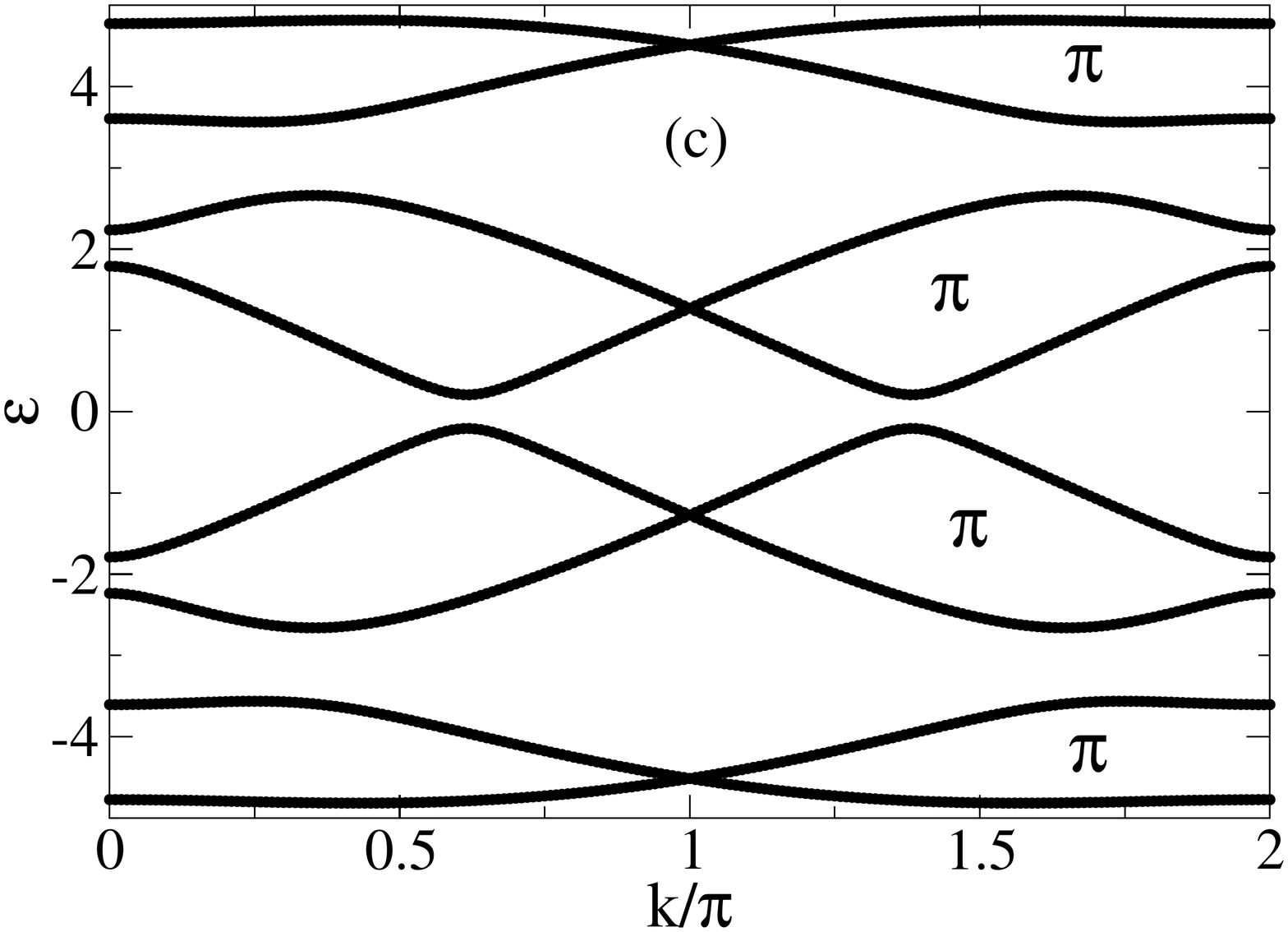}
\includegraphics[width=0.34\textwidth]{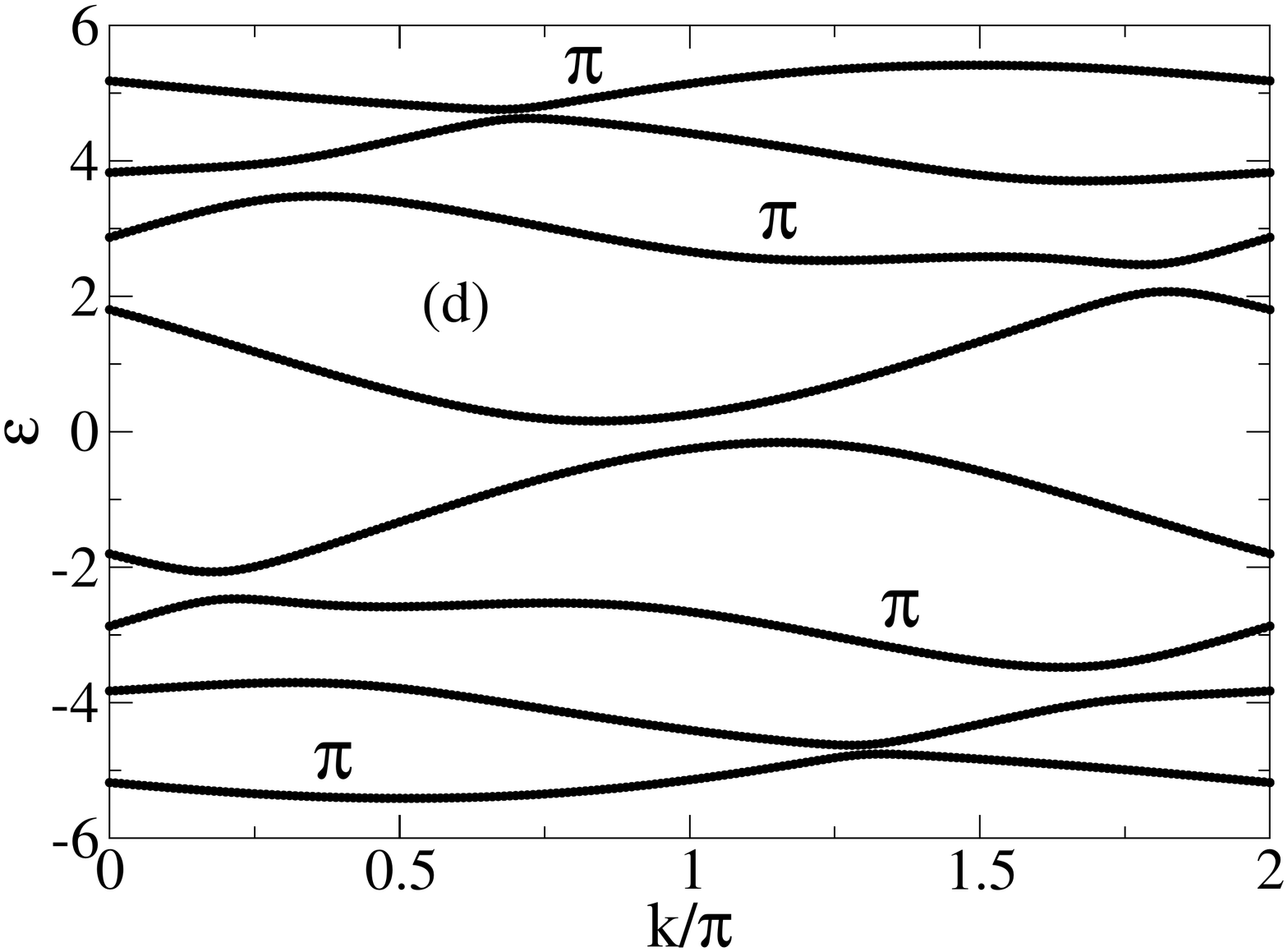}
\caption{\label{berryk}
Energy bands for 
a) $\lambda=2$ and no external magnetic field, 
b) no spin-orbit coupling  and $B_x=1, B_y=0, B_z=1.5$, 
c) $\lambda=2$ and $B_x=1, B_y=B_z=0$ and d) with spin-orbit coupling
$\lambda=2$ and $B_x=1, B_y=0, B_z=1.5$.
In panels c) and d) we show the bands with non-trivial Berry phase, $\pi$ phase.
In panel d) the bands are all separated and the Berry phase is for each band
individually. The bands have either a $\pi$ Berry phase or a Berry phase of zero.
In panel c), since the bands have degenerate points, the Berry phases indicated are for
groups of two bands. In panels a) and b) the Berry phases are not quantized and have
arbitrary real values.
}
\end{figure}

\subsection{Berry phase}

In the case of bands with a finite gap between them, the Berry phase may be easily determined.
In general, a Berry phase does not have a topological nature. 
However, 
the presence of some symmetries, like inversion, chiral or charge conjugation symmetries \cite{zak,benalcazar},
ensures a quantization of the Berry phase.
This is calculated $mod(2 \pi)$ and a trivial band has zero Berry phase and a topological band
has a Berry phase of $\pi$. 
The Berry phase of a given band may be calculated in a standard way \cite{niu,resta,fukui,benalcazar}. One divides the
Brillouin zone considering $N$ discrete points, $k \in [0,2\pi]$, with momenta taking the values $k_1, \cdots ,k_N$.
The band's Berry phase $\gamma_B$ may be obtained defining the link variable $U(k_l)=\varphi^*(k_l)
\varphi(k_{l+1})$, and summing over $k_l$ as
\be
\gamma_B=-i \sum_{l} \log U(k_l)
\ee
where $\varphi(k_l)$ are the eigenstates of the Hamiltonian in momentum space. 

This procedure can be generalized if there are degenerate points between the
bands taking a link variable as $U(k_l)= det \mathbf{U}(k_l)$, involving the determinant of the
matrix
\be
\mathbf{U}_{ij}(k_l) = \varphi_i^* (k_l) \varphi_j(k_{l+1})
\ee
where $1\leq i,j \leq N$ run over the eigenstates and $\varphi_i(k_l)$ is the eigenstate at momentum
$k_l$ of the $i$th band.

In Fig. \ref{berryk} we show the energy bands and non-trivial (quantized) Berry phases, considering
a unit cell with four sites, $\gamma=4$.
With no spin-orbit coupling (zero electric field) or zero external magnetic
field, the bands are gapless (as shown in Fig. \ref{berryk}(a),(b) ) and the Berry phases are not quantized.
Combining the two effects, in some regimes gaps appear between the various
bands. The charge-conjugation symmetry implies that there are pairs of states 
that are related by $\epsilon_n(k) \rightarrow -\epsilon_m(-k)$, where $n,m$ are two
bands.
In Fig. \ref{berryk}(c) we consider the addition of a magnetic field along the $x$ direction.
Some gaps appear and the bands appear in groups of two. The sum of the Berry phases
of the two lowest bands adds to $\pi$ as well as the sum of all following pairs of bands.
We show in Fig. \ref{berryk}(d) an example of a situation where the various bands are separated
by finite gaps as a result of adding a magnetic field $B_z$ different from zero. 
The results for the Berry phases show that some bands are trivial and four bands
are non-trivial, with a Berry phase $\gamma_B=\pi$. 
Note that the Berry phases of all the bands below zero energy add to zero, as a consequence
of the charge-conjugation symmetry.
Also, even though the model is in class $C$, which is not topological in the usual scheme, 
the Berry phases of individual bands show quantization properties.

\begin{figure*}
\centering
\includegraphics[width=0.4\textwidth]{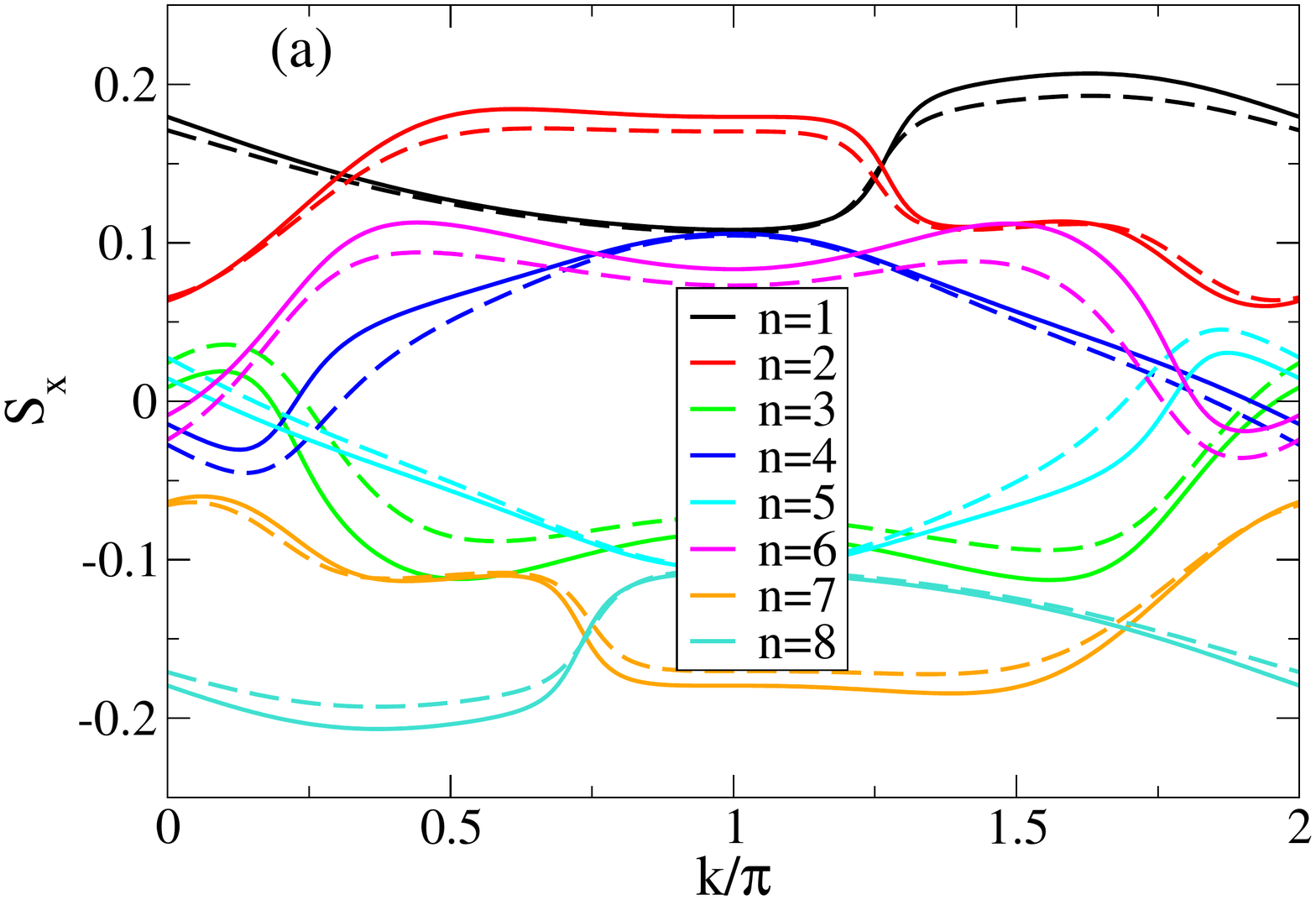}
\includegraphics[width=0.4\textwidth]{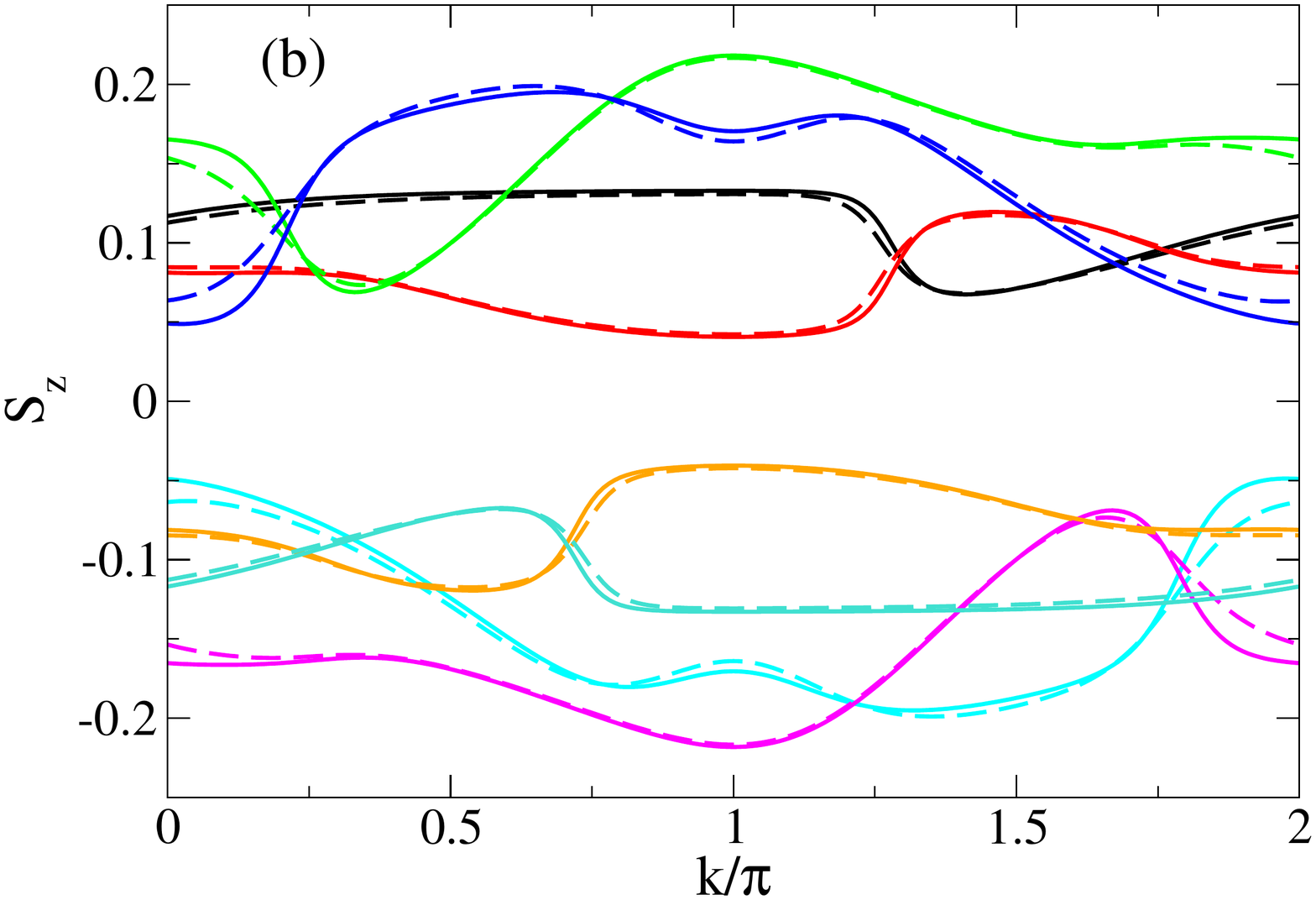}
\caption{\label{spink}
Spin density averages $S_x$ and $S_z$, as a function of momentum, for each band
for $\gamma=4$, for a trivial case (dashed lines) and a case with quantized Berry phase (solid lines).
The case of quantized Berry phase has parameters $\lambda=2, B_x=1.2, B_y=0, B_z=1.5$ and the trivial
case has the same parameters except $B_y=0.5$.
}
\end{figure*}

As mentioned above, the quantization of Berry phase may result from inversion symmetry,
chiral symmetry or charge-conjugation symmetry \cite{benalcazar}. 
The helix structure and the presence of a magnetic field along $x$ and $z$ directions implies no inversion
symmetry. In the presence of a magnetic field, $\mathbf{B} \neq 0$ there is no chiral symmetry. 
However, there is a hidden symmetry if $B_y=0$ (for $\gamma=4$).
It is of the type
$O h(k) O^{-1} = - h(-k)$,
and may se seen as a combination of inversion symmetry
$I h(k) I^{-1} = h(-k)$
and chiral symmetry
$\Pi h(k) \Pi^{-1} = -h(k)$
specifically $O=\sigma_2 \otimes \sigma_1 \otimes \sigma_2$.
The various symmetries chiral symmetry, hidden symmetry and charge-conjugation symmetry
also only hold if the chemical potential vanishes and the number of occupied bands equals
the number of unnoccupied bands. 
There is a relation between the Zak phase, the electric polarization (dipole moment) and Wilson
loops \cite{vanderbilt,resta2,niu,benalcazar}. Using the symmetry properties of the Wilson loops
and its relation to the dipole moment, $p$, \cite{benalcazar} as the sum of the eigenvalues of the Wilson loop
via the Wannier centers (phases of the eigenvalues of the Wilson loop), it can be shown that, in the
presence of inversion symmetry, the dipole moment (normalized to $1$) of the states summed over the occupied
bands, satisfies $p_{occ}=-p_{occ}$ implying that it is quantized to $0,1/2$, which leads to a
Zak phase of $0,\pi$. A similar analysis if there is a chiral symmetry leads to the
result that $p_{occ}=p_{unnocc}$, since the chiral symmetry connects states with positive energy
with states with negative energy. Using the general result that $p_{occ}+p_{unnocc}=0$ (summing over all
bands topology is trivial) we get the same result that $p_{occ}=-p_{occ}$ and, therefore, it is
quantized. A similar result may be shown for charge-conjugation symmetry \cite{benalcazar}.
In the case of the hidden symmetry we get that $p_{occ}=-p_{unnocc}$ since this symmetry changes the
sign of the momentum and relates the positive energy states with the negative energy states. So, at first sight,
no conclusion about the quantization may be obtained. 
At zero chemical potential the charge-conjugation symmetry is present (even in the presence of a magnetic
field) and there is quantization of the Berry phase, considering the sum over the occupied bands.
Indeed, we find, by explicit calculation of the Berry phases, that the sum over the lower four bands always
leads to a zero Berry phase. It is quantized but topologically trivial. This is consistent with the
class $C$, which is not topological in $1d$.
As mentioned above, if $B_y=0$ we find that some of the bands, or groups of bands, show a quantized $\pi$ Berry phase.
Adding $B_y \neq 0$ all the bands show non-quantized Berry phases, but are such that the sum over the occupied
bands still yields a quantized, vanishing Berry phase, as imposed by the charge-conjugation symmetry.
Even though the charge-conjugation and the hidden symmetry only hold if the chemical potential vanishes,
it turns out that a finite chemical potential placed in a gap between the bands leads to a Berry phase of
each band that is the same as for $\mu=0$. As a consequence, it is possible to select the chemical potential such that
the sum of the Berry phases over the set of occupied bands is topologically non-trivial and quantized to $\pi$.
This only takes place if $B_y=0$, and therefore we argue that    
the quantization of the Berry phase is associated with the hidden symmetry, since the charge-conjugation
symmetry implies $p_{occ}=p_{unnocc}$ and the hidden symmetry implies $p_{occ}=-p_{unnocc}$. 

The non-trivial Berry phase found for $\gamma=4$ can be generalized to any number of points in the unit
cell. Recalling that in the rest frame of an electron the effect of the electric field can be seen
as an effective magnetic field, $\mathbf{b}$, that couples to the electron spin via a Zeeman coupling,
we can see that the effective field has components $\mathbf{b}\sim (-E_y,E_x,0)$. We can see that at site $j$
of the unit cell, $b_y/b_x=-1/\tan [\Phi \left(j-1 \right)]$, with $\Phi=2\pi/\gamma$. 
We have found that the condition for a non-trivial
Berry phase can be obtained selecting
\be
\frac{B_y}{B_x}= \frac{1}{\tan \Phi}
\label{quantization}
\ee


In a more general case, for which
\be
\frac{E_y}{E_x} = \tan \left( \Phi + \delta_1 \right)
\ee
with $\delta_1$ any phase,
we find that the Berry phase is quantized if we
take
\be
\frac{B_y}{B_x} = \frac{1}{\tan \left( \Phi + \delta_2 \right)}
\ee
with $\delta_2=-\delta_1 \pm l \pi$, with $l$ an integer.

We also find regimes of quantized Berry phases for a model of spinless electrons
but with a non-trivial hopping structure, due to the presence of more than one orbital state, and considering
both nearest neighbor and next-nearest neighbor terms (with a complex part) and therefore
an external magnetic field is not required.
This is shown in Appendix A.

\begin{figure}
\centering
\includegraphics[width=0.41\textwidth]{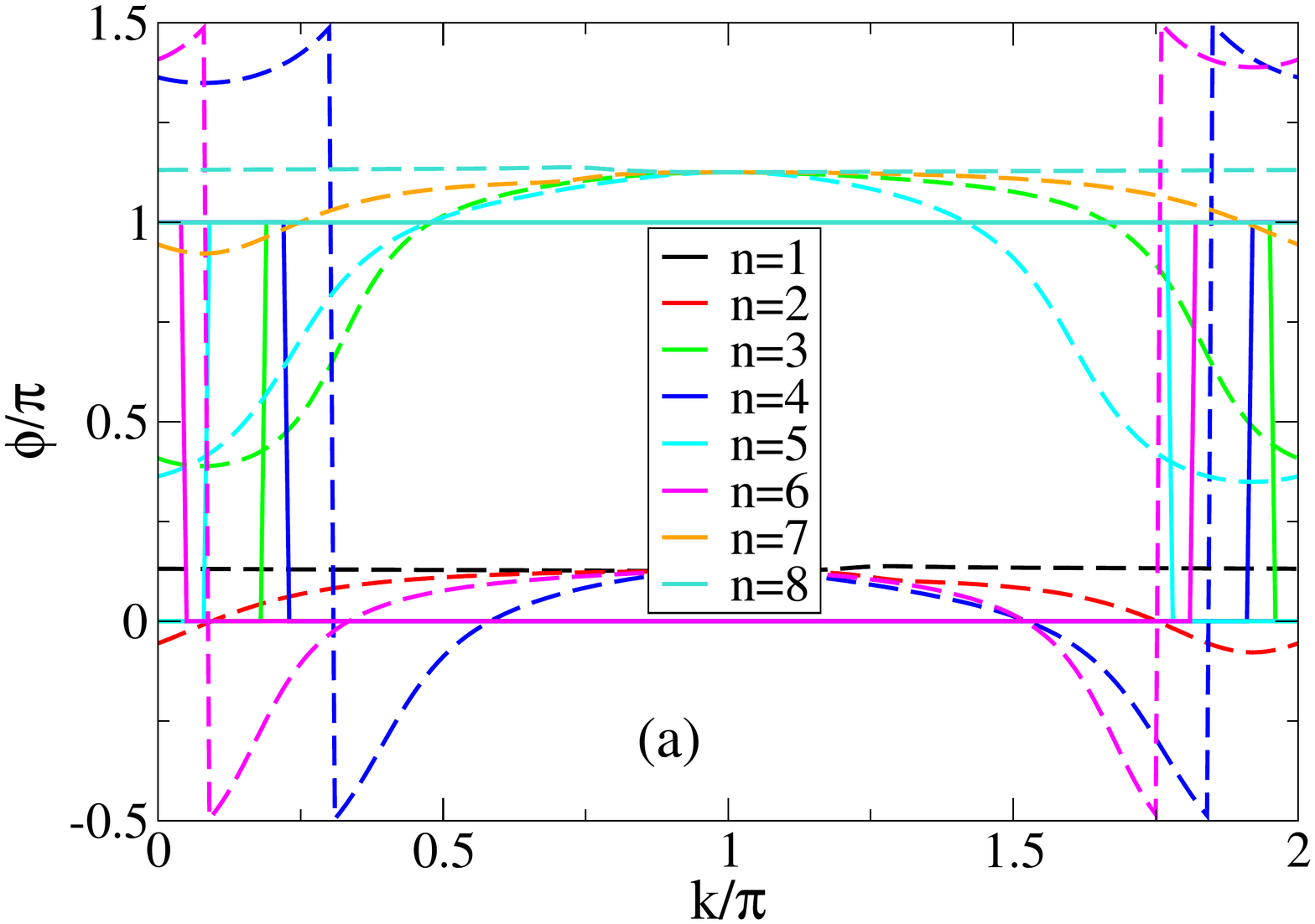}
\includegraphics[width=0.41\textwidth]{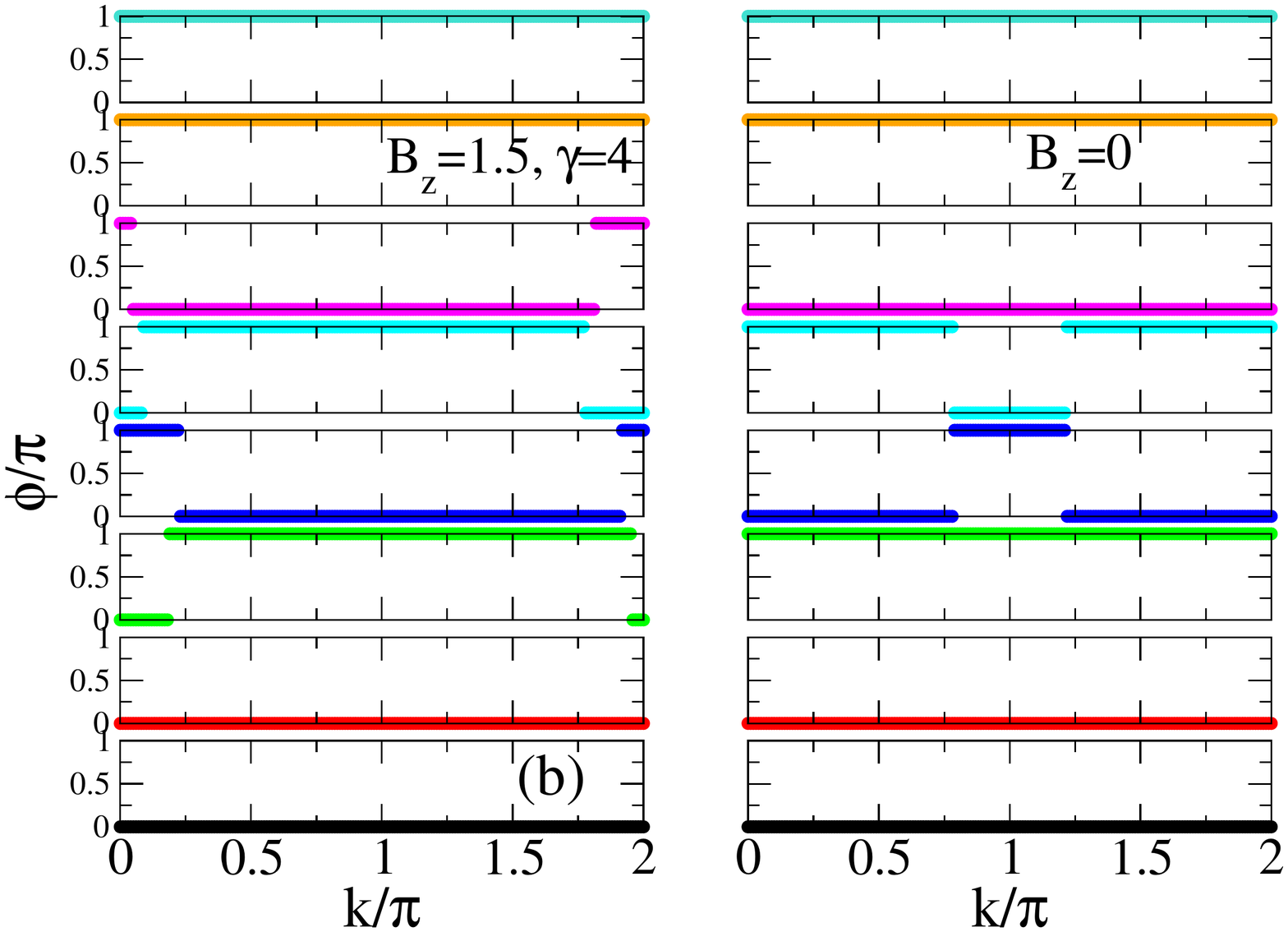}
\includegraphics[width=0.41\textwidth]{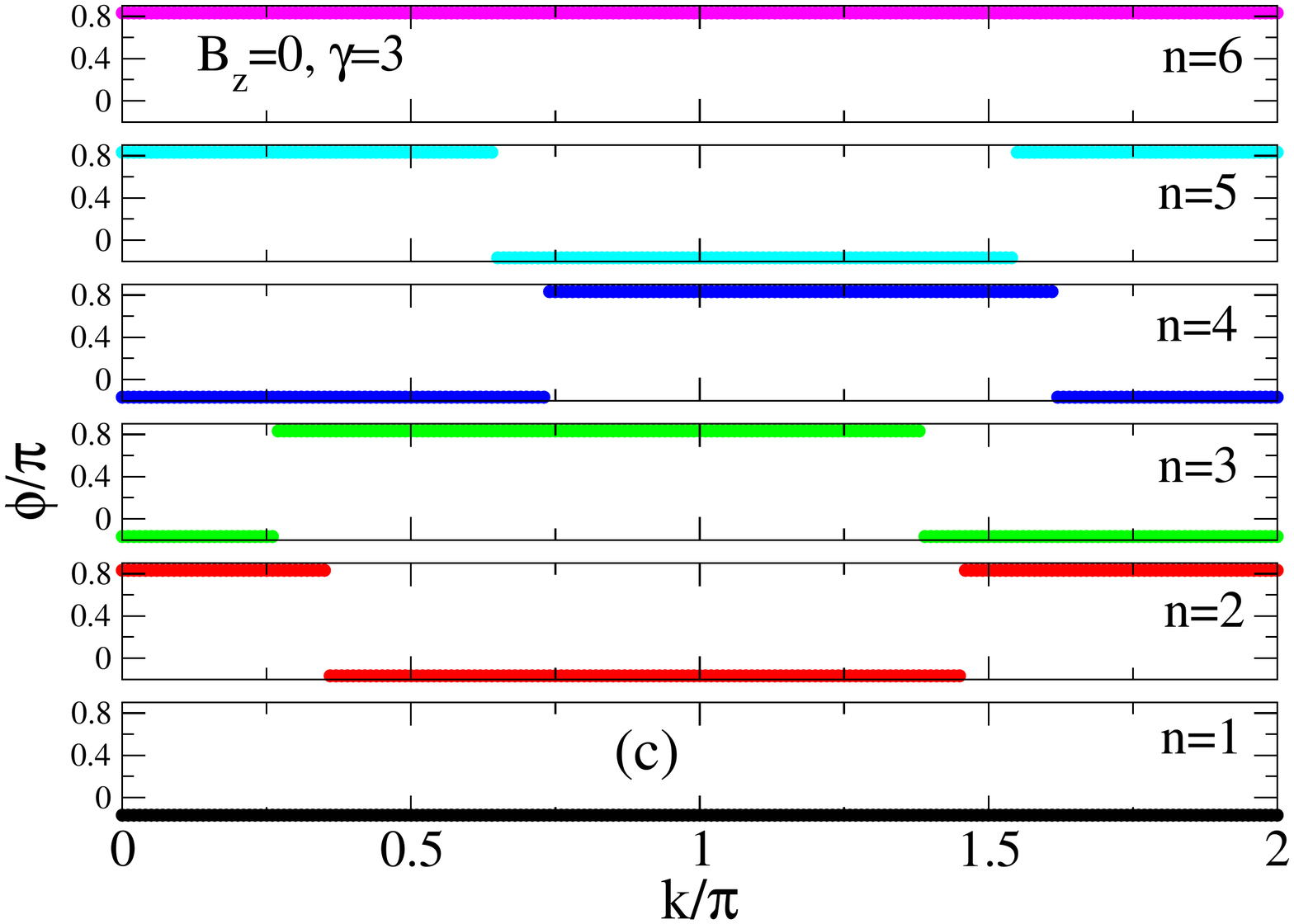}
\caption{\label{polark}
a) Comparison of polar angle of spin density average over an eigenstate of momentum $k$ and band $n$, 
between a trivial case (dashed lines) and
a non-trivial (quantized Berry phase) case (solid lines), as a function of momentum, for the various bands.
The parameters are the same as in Fig. \ref{spink}.
b) Comparison for $\gamma=4$ of the polar angles for the various bands for $B_z=1.5$ and
$B_z=0$.
c) Polar angle for $\gamma=3$ for $B_z=0$. 
}
\end{figure}

\section{Momentum space spin density and polar angle quantization}

The spin density operator components may be defined by
\be
S_{\beta}(k) = \frac{1}{\gamma} \sum_{\alpha=1}^{\gamma} \sum_{m_s,m_s^{\prime}} c^{\dagger}(k,\alpha, m_s)
\sigma_{m_s,m_s^{\prime}}^{\beta} c(k,\alpha,m_s^{\prime})
\ee
where $\beta=x,y,z$ and $\sigma^{\beta}$ are Pauli matrices. We may  consider the average value of these operators
in the eigenstates of the Hamiltonian, for a given band and momentum value.
That is, we consider for a given momentum value and energy band, $n$,
\bea
\langle S_x(k,n) \rangle &=& \frac{1}{\gamma} \sum_{\alpha=1}^{\gamma} 
\left( u_n^*(k,\alpha,\uparrow) u_n(k,\alpha,\downarrow) \right.
\nonumber \\
&+& \left. u_n^*(k,\alpha,\downarrow) u_n(k,\alpha,\uparrow) \right)
\nonumber \\
\langle S_y(k,n) \rangle &=& \frac{1}{\gamma} \sum_{\alpha=1}^{\gamma} 
\left( -i u_n^*(k,\alpha,\uparrow) u_n(k,\alpha,\downarrow) \right.
\nonumber \\
&+& \left. i u_n^*(k,\alpha,\downarrow) u_n(k,\alpha,\uparrow) \right)
\nonumber \\
\langle S_z(k,n) \rangle &=& \frac{1}{\gamma} \sum_{\alpha=1}^{\gamma} 
\left( |u_n(k,\alpha,\uparrow)|^2
-
|u_n(k,\alpha,\downarrow)|^2 \right)
\nonumber \\
\eea
where $u_n(k,\alpha,m_s)$ is the eigenfunction of state $n,k,\alpha,m_s$.
We may also define, for each momentum value and energy band, the normalized spin vector
\be
\hat{S}_n(k)=\frac{\left( \langle S_x(k,n) \rangle ,\langle S_y(k,n) \rangle ,
\langle S_z(k,n) \rangle  \right)}{\sqrt{ \langle S_x(k,n) \rangle^2+
\langle S_y(k,n) \rangle^2+ \langle S_z(k,n) \rangle^2}}
\ee
with a direction in space parametrized by the spherical coordinate angles $(\theta(k,n), \phi(k,n))$.
We will show that when the Berry phase is quantized, the polar angle gets
quantized as a function of momentum.

In Fig. \ref{spink} we show the results for the spin density average as a function of momentum for the various
bands for a unit cell of four sites. We consider the spin densities along the $x$ and $z$ directions.
Also, we compare the results for a system which shows a quantized Berry phase and one which is
not quantized. The results are in general similar, since the trivial regime 
is obtained adding
a small magnetic field along the $y$ direction. However, if we consider the polar angle for the same
two cases, we show in Fig. \ref{polark} that in the quantized Berry phase regime, the polar
angle also becomes quantized, while in the trivial case it is in general not quantized.
This result holds for any unit cell size, $\gamma$. The angle with respect to the $z$ axis shows no quantization.

\begin{table*}
\label{Tab:table}
  \caption{Quantized Berry phase, $\gamma_B$, and polar angle, $\phi(k,n)$, for the cases of $\gamma=3,4,5,6$, when
the conditions of eq. (\ref{quantization}) are met.
   Here $k$ is the momentum and $n$ is the band index, $n=1,\cdots ,2 \gamma$.}
  \begin{center}
    \label{tab:table1}
    \begin{tabular}{c c c c c c c c c c c c}
 $\gamma=3$ & & & $\gamma=4$ & & &
 $\gamma=5$ & & & $\gamma=6$ & & \\
\hline
n & $\gamma_B$ & $\phi(k,n)/\pi$ & n & $\gamma_B$ & $\phi(k,n)/\pi$ &
n & $\gamma_B$ & $\phi(k,n)/\pi$ & n & $\gamma_B$ & $\phi(k,n)/\pi$  \\
\hline
1 & $\pi$ & -1/6 & 1 & $\pi$ & 0 &
1 & $\pi$ & 1/10 & 1  & $\pi$ & 1/6 \\
2 & 0 & 5/6,-1/6 & 2 & 0 & 0 &
2 & 0 & 1/10 & 2 & 0 & 1/6 \\
3 & $\pi$ & -1/6,5/6 & 3 & $\pi$ & 0, 1 &
3 & 0 & 1/10,11/10 & 3 & $\pi$  & 1/6  \\
4 & $\pi$ & -1/6,5/6 & 4 & 0 & 1,0 &
4 & $\pi$ & 11/10,1/10 & 4 & 0 & 7/6,1/6 \\
5 & 0 & 5/6,-1/6 & 5 & 0 & 0,1 &
5 & 0 & 1/10,11/10 & 5 & 0 & 1/6,7/6 \\
6 & $\pi$ & 5/6 & 6 & $\pi$ & 1,0 &
6 & 0 & 1/10,11/10 & 6 & 0 & 1/6,7/6 \\
 &  &  & 7 & 0 & 1 &
7 & $\pi$ & 11/10,1/10 & 7 & 0 & 7/6,1/6 \\
 &  &  & 8 & $\pi$ & 1 &
8 & 0 & 1/10,11/10  & 8 & 0 & 7/6,1/6 \\
 & & & & & &
9 & 0 & 11/10 & 9 & 0 & 1/6,7/6 \\
& & & & & &
10 & $\pi$ & 11/10 & 10  & $\pi$  & 7/6 \\
& & & & & & 
 &  &  & 11 & 0 & 7/6 \\
& & & & & &
 &  &  & 12 & $\pi$ & 7/6 \\
\end{tabular}
\end{center}
\end{table*}

In Fig. \ref{spink} and in Fig. \ref{polark}(a) we consider 
the case where $B_z \neq 0$. This was chosen earlier
since applying a magnetic field in the parallel and transverse directions, with respect to the
chiral axis, allows a simplified problem where all bands are separated by gaps. As shown in 
Fig. \ref{berryk}(c),
turning off $B_z$ leads to the closing of some gaps, even though a quantized Berry phase is still
found for groups of two bands. Turning off $B_z$ 
we find that the polar angle is also quantized.
In Fig. \ref{polark}(b) we compare the cases of $B_z=1.5$ and $B_z=0$, for $\gamma=4$, which shows
the quantization in both cases.  In Fig. \ref{polark}(c) we show
the quantization of the polar angle for $\gamma=3$ when $B_z=0$. The case of vanishing $B_z$ will
be considered later in the spin transport across the helix. 

We summarize the results for the Berry phase, $\gamma_B$, and the polar angle,
$\phi(k,n)/\pi$, for each band, $n$, in Table I for 
$\gamma=3,4,5,6$ and $B_z=1.5$.
We see that for the lowest energy levels the spin density is aligned along $\mathbf{B}$ while for the higher levels
the spin density is aligned along the $\mathbf{B}+\pi$ direction. In the intermediate levels the spin density
is aligned in one or the other direction depending on the momentum value, $k$. 
There is a competition between
the external magnetic field and the effective field that is the result
of the electric field of the helix. In the non-quantized cases this leads in
general to a polar angle that varies with the momentum and band, while
in the quantized Berry phase cases the polar angle has plateaus. In some cases
the competition between the two magnetic fields leads to a situation
where for a given band there is only one plateau. This happens for the
lowest and/or the highest energy bands.

\begin{figure}
\centering
\includegraphics[width=0.4\textwidth]{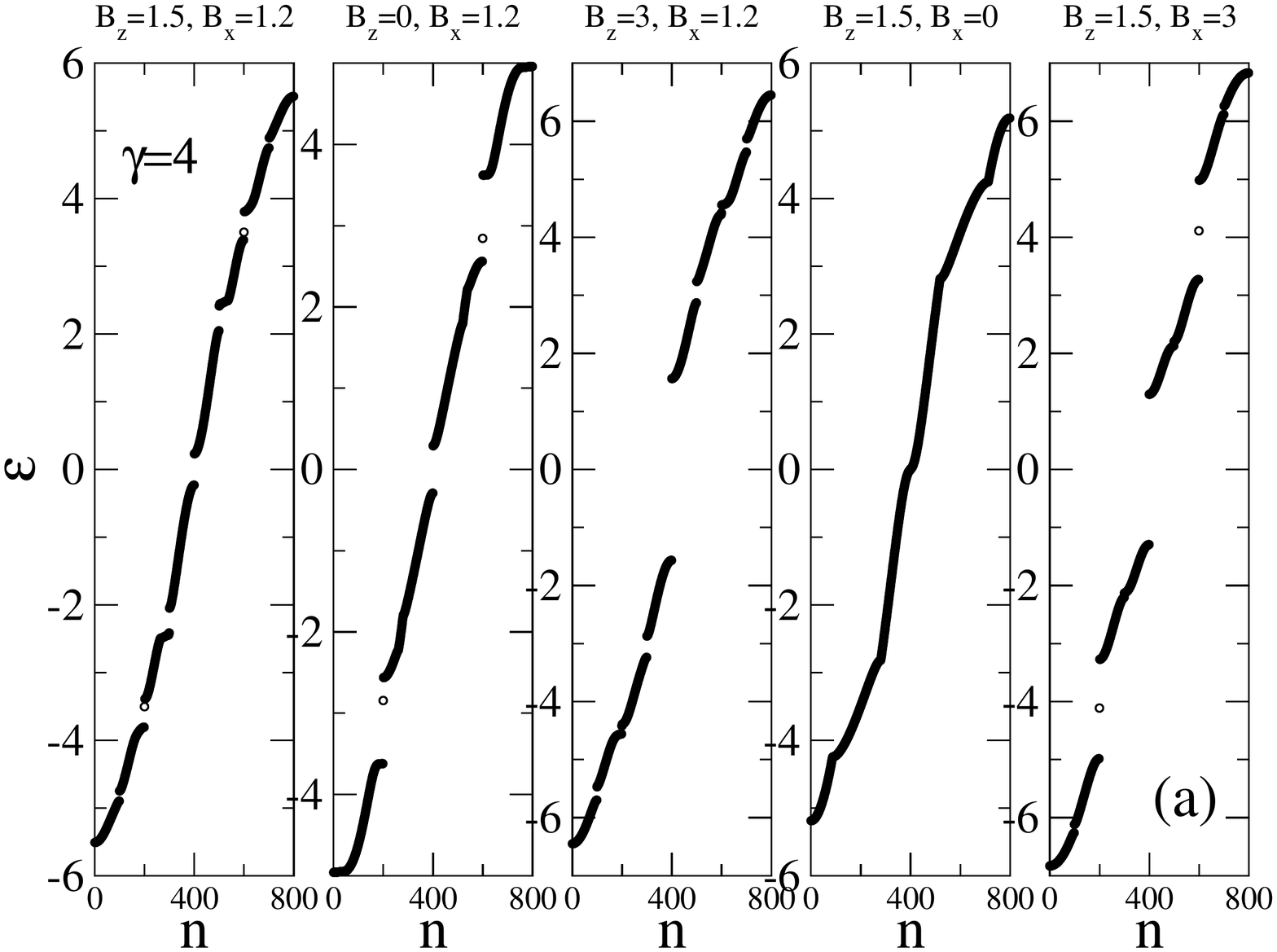}
\includegraphics[width=0.4\textwidth]{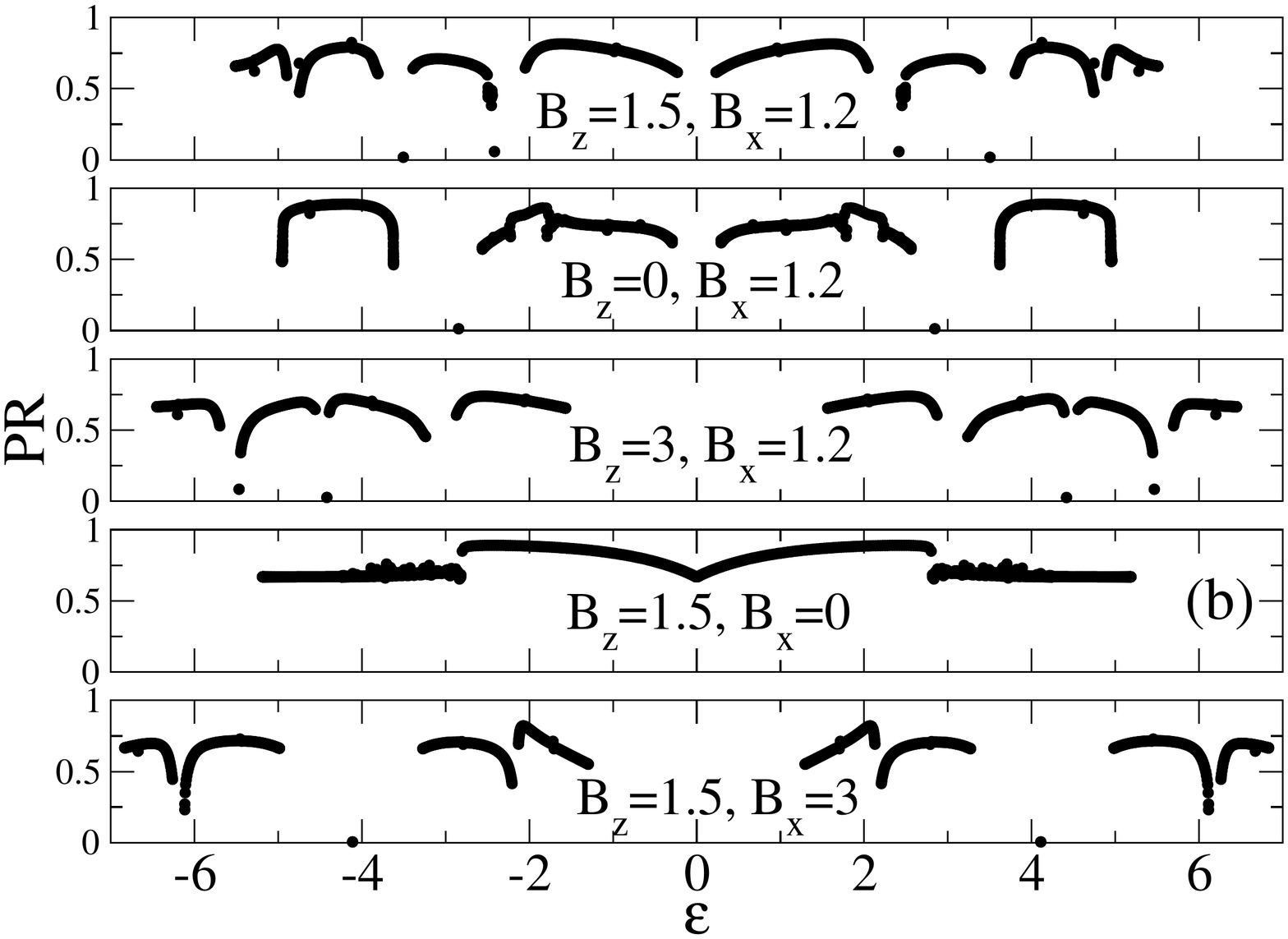}
\includegraphics[width=0.4\textwidth]{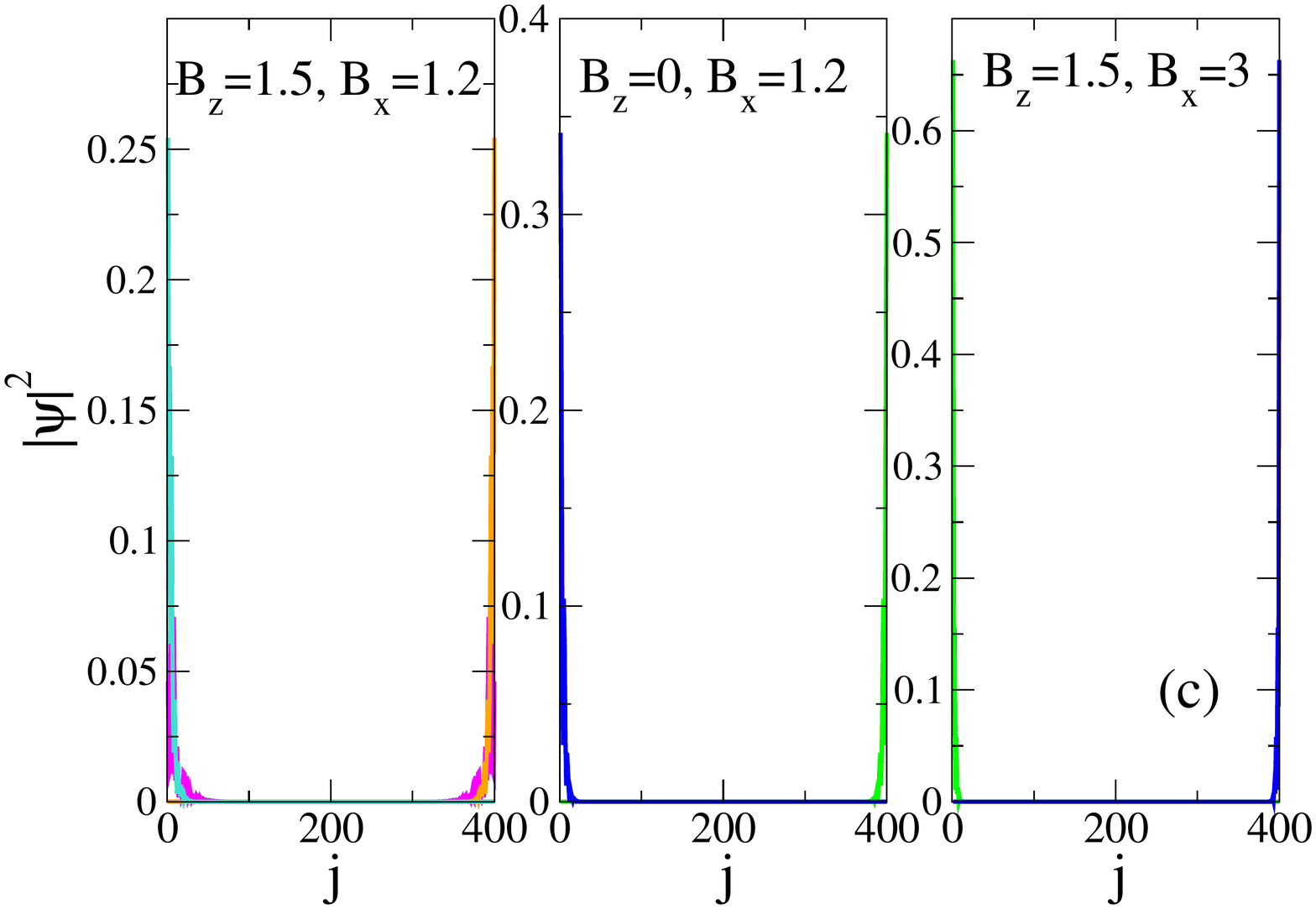}
\caption{\label{rspin}
Real space diagonalization with open boundary conditions for a finite system of 100 unit cells.
a) Comparison of energy values for $\gamma=4$ and different values of the external magnetic
field, where $n$ is the level label.
b) Corresponding results for the participation ratio of each energy state.
c) Absolute value of the eigenfunctions of the edge states as a function of space location, $j$, for
some values of the magnetic field. 
}
\end{figure}

\section{Real space description and edge states}

\begin{figure}
\centering
\includegraphics[width=0.4\textwidth]{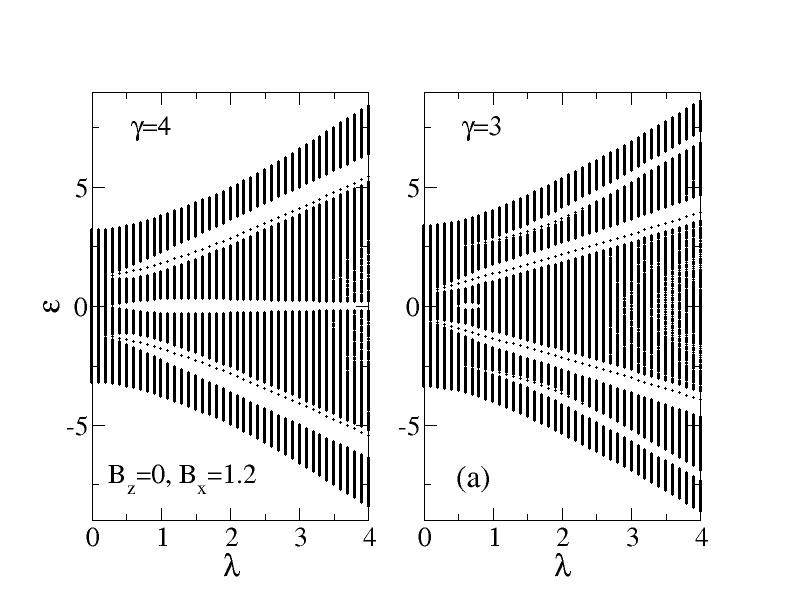}
\includegraphics[width=0.4\textwidth]{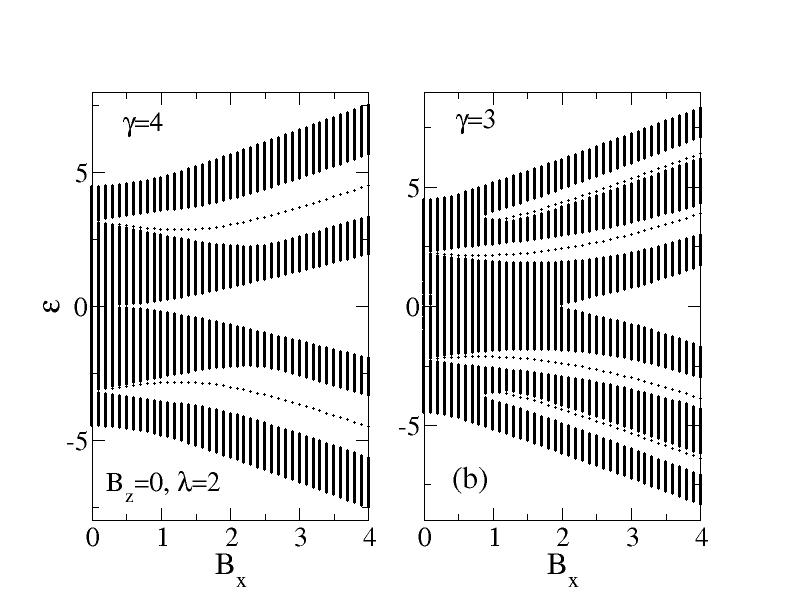}
\includegraphics[width=0.4\textwidth]{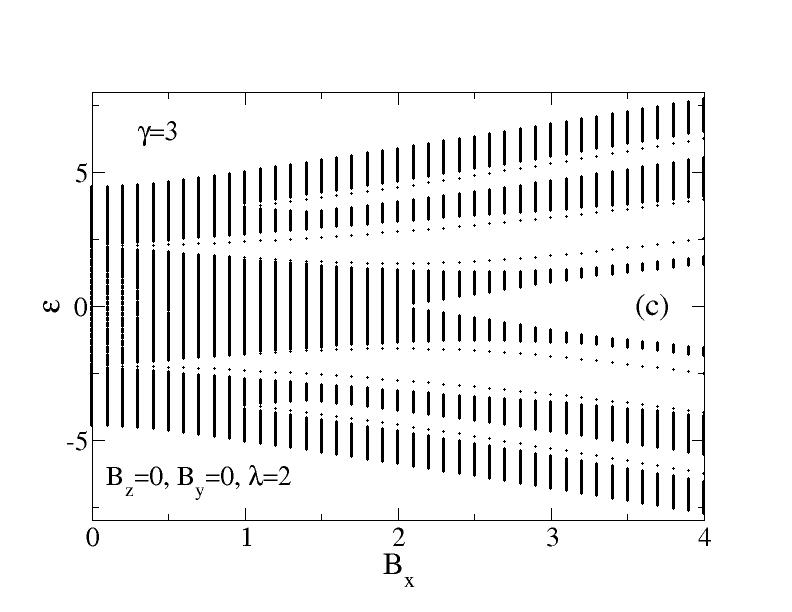}
\caption{\label{rspin2}
Comparison of energy values for $\gamma=4$ and $\gamma=3$ as a function
of a) $\lambda$ and b) $B_x$, for the quantized Berry phase and c) for $\gamma=3$
with $B_y=0$ (non-quantized case).
}
\end{figure}

The presence in some regimes of a quantized Berry phase may suggest the presence of
edge states at the ends of the helix. These are revealed diagonalizing the Hamiltonian
in real space and considering open boundary conditions. The results are shown in Fig. \ref{rspin}.
In the top panel we show the energy levels for similar parameters as those taken in Fig. \ref{berryk}
with $\lambda=2$ and $B_x=1.2, B_y=0, B_z=1.5$, together for other values of the magnetic
field. There are states inside the gaps, but there are no zero energy states.
In non-chiral multiband systems the Zeeman term has a similar effect \cite{ssh2,bahari}.
Recall that the sum of the Berry phases of the bands below
zero energy is zero, consistently with no edge states in the gap surrounding zero energy.
The states inside the gaps are well localized, as shown
in Fig. \ref{rspin}(b) by the participation ratio of each level and by the wave functions shown
in Fig. \ref{rspin}(c), that are well localized at the edges of the chain.
The participation ratio for a given state is defined as 
\be
PR = \frac{L^{-1}}{\sum_j \left( |\psi_j|^2 \right)^2}
\ee
where $|\psi_j|^2$ includes the sum over the two spin components at site $j$.
It is such that is of the order of one if the state is extended and is of the order of
$1/L$, where $L$ is the system size, if the state is localized. 
We may argue that the edge states appear in cases where the sum of the Berry phases below a given gap
is non-trivial (this is particularly clear in the case that $B_z=0$),
as would be expected, even though the model is in class
$C$. 
Even though this picture seems appealing, it turns out that the edge states remain in some
cases, even when the Berry phase is not quantized to zero or $\pi$.
The momentum space band structure and the energy states for the finite
system agree qualitatively, but the boundary conditions used are
not the same. Also, unless the indirect gaps between the bands
are positive, the possible
edge (localized) states merge into the continuum, are not seen and become extended, as revealed for
instance by the absence of small PRs.

\begin{figure}
\centering
\includegraphics[width=0.4\textwidth]{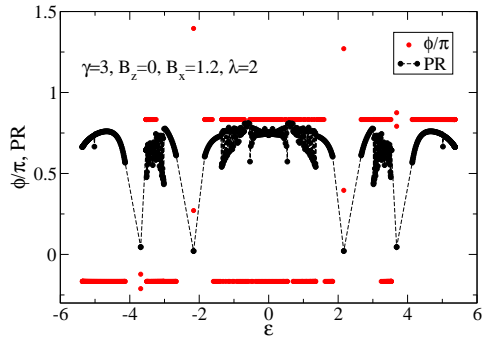}
\caption{\label{rpolar}
Participation ratio and spin polar angle for the energy states of a chain with unit cell with
$\gamma=3$, with a finite size of 100 cells and using open boundary conditions, for
$B_z=0, B_x=1.2, \lambda=2$ in the topological case with $B_y/B_x=1/\tan \Phi$. The polar angle
of the bulk states is quantized along the direction of the magnetic field in the transverse plane and along
the opposite direction. Note that the polar angle is not quantized when we consider the edge states,
for which the participation ratio has large drops.
}
\end{figure}

In the top two panels of Fig. \ref{rspin2} we show the dispersion of the localized 
edge states as a function of $\lambda$ and
$B_x$ for $\gamma=3,4$, in regimes where the Berry phase is quantized. 
In Fig. \ref{rspin2}(c) we consider a similar plot of the energy levels for $\gamma=3, B_z=0, \lambda=2$
but taking $B_y=0$, which can be shown in momentum space to lead to a phase with
a non-quantized Berry phase and a non-quantized spin polar angle. But, as clearly shown, edge states are found
inside the various gaps, and these states are well localized at the edges of the system.

As in the case where the spin operators were averaged over the momentum space eigenstates, we also find for the
bulk states of the finite system a quantization of the spin polar angle (in the topological regimes).
This contrasts
with the result that the edge states do not display this quantization.
This is shown in Fig. \ref{rpolar} where the polar angle of each eigenstate is shown, 
together with the participation
ratios, for $\gamma=3$. While the bulk states display the expected quantization, the edge states, identified
by the small participation ratios, do not lead to a quantized polar angle.

\section{Generalized models: lattice distortion and Rashba spin-orbit}

It is interesting to consider generalizations of the model.
We may consider that along the helix some dimerization, trimerization, etc, may
occur, that leads to a non-homogeneous set of hopping terms between the various neighbors.
For instance, a unit cell with
two sites, $\gamma=2$, which may be dimerized, or a unit cell with $\gamma=3$ which may be trimerized in some
way. The first case is similar to a SSH model (with spin added). The model by itself has a topological
regime, with zero energy edge states, quantized winding number and quantized Berry phase.

For instance, taking $\gamma=2$, if the two hoppings $t_1,t_2$ are not equal,
topology emerges if $t_2>t_1$. Including spin in a SSH model, the bands become degenerate, but considering
a magnetic field along the direction of the chain, $B_z \neq 0$ lifts the degeneracy and one finds the
expected topological bands with $\pi$ Berry phase. Adding the spin-orbit term that results from
the chiral structure, the topology
is maintained, and due to the spin-orbit coupling a transverse spin density along the $y$ direction is found. 
This leads naturally to a quantization of the polar angle along either $\phi=\pi/2$ or $3 \pi/2$.

\begin{figure}
\centering
\includegraphics[width=0.4\textwidth]{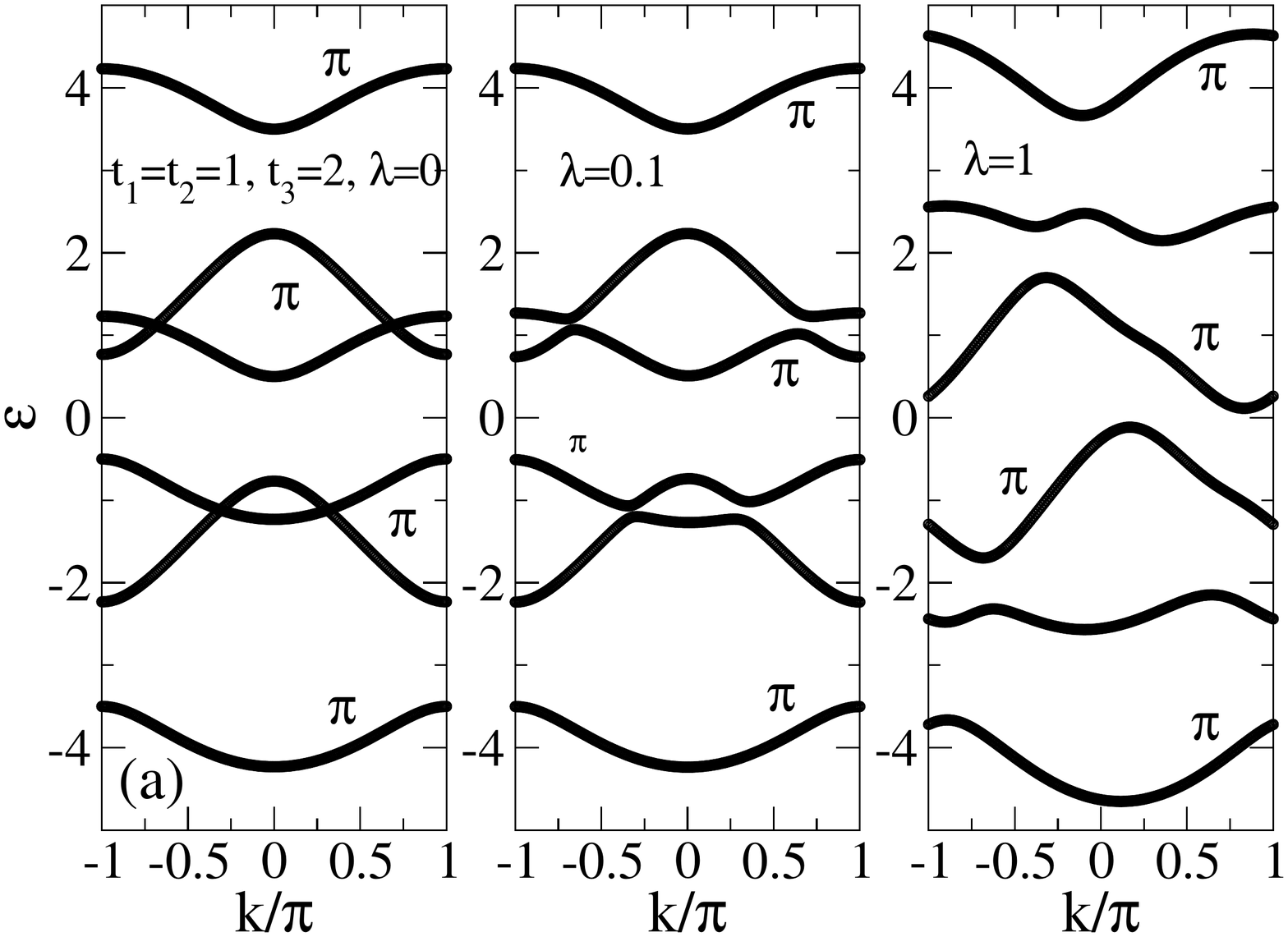}
\includegraphics[width=0.4\textwidth]{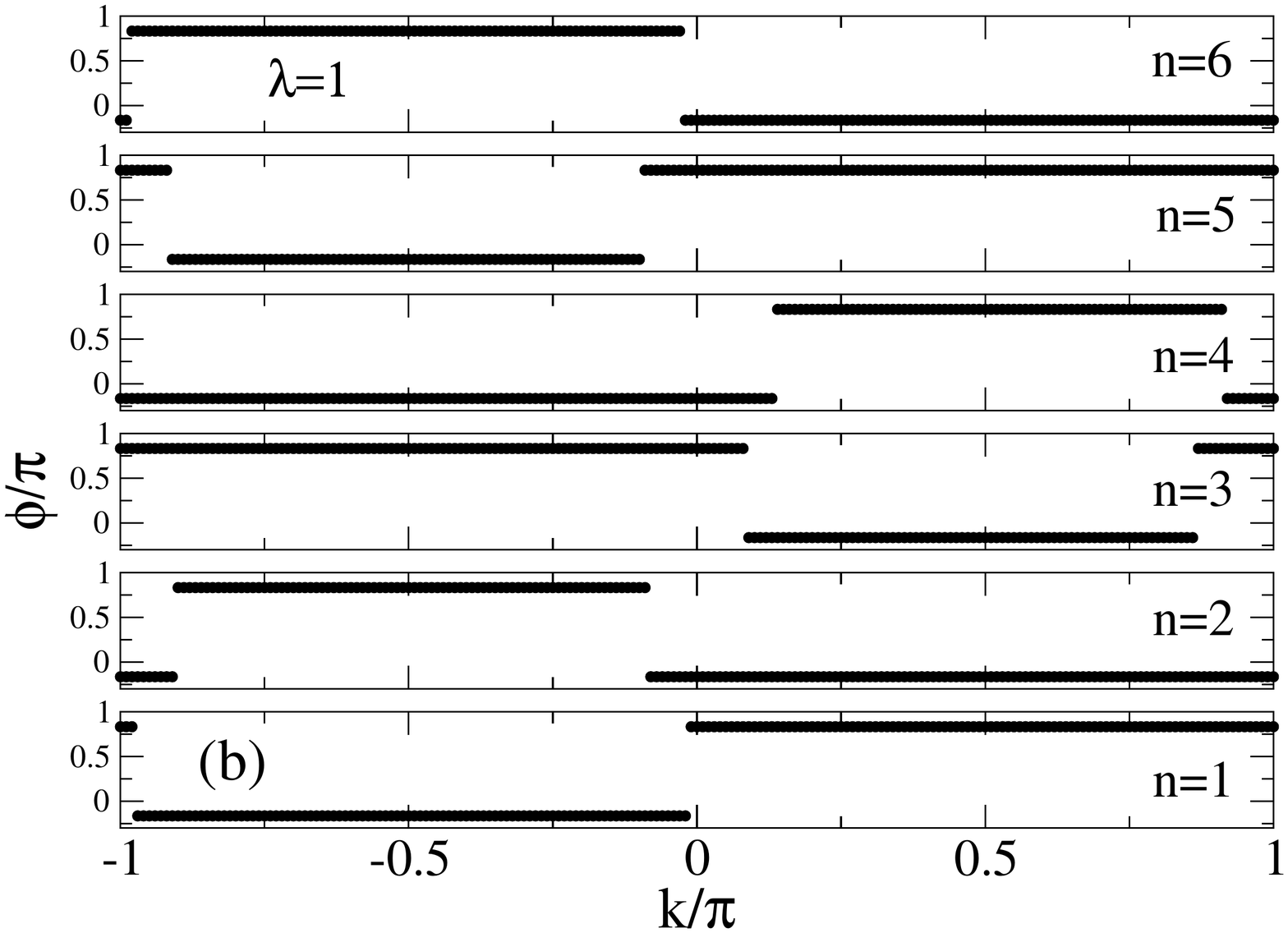}
\caption{\label{trimer}
a) Energy bands of a chain with $\gamma=3$ and non-homogeneous hoppings, with
$t_1=t_2=1, t_3=2$, for different values of spin-orbit coupling $\lambda=0,0.1,1$.
b) Quantization of the polar angle as a function of momentum, for topological case, for each band, for
$\gamma=3$ and $\lambda=1$. 
}
\end{figure}

\begin{figure}
\centering
\includegraphics[width=0.4\textwidth]{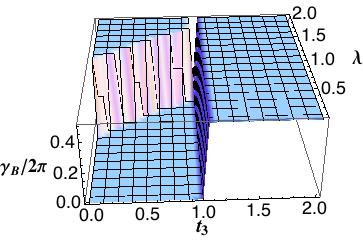}
\caption{\label{t3lbd}
Berry phase of the combined two lowest energy levels, as a function of $t_3,\lambda$, for $B_x=B_y=0, B_z=1.5$, for $t_1=t_2=1$.
The Berry phase of the lowest level is non-trivial for $t_3>1$ and the Berry phase of the second level is non-trivial
when $t_3<1$ and $\lambda$ is large enough.
}
\end{figure}

Taking $\gamma=3$ and selecting the three hoppings such that $t_3>t_1,t_2$, a
non-trivial quantized spin polar angle is found. Unexpectedly, for $t_3<t_1,t_2$ a quantized polar angle is also
found, even though the Berry phase vanishes (as expected for a trivial regime of a trimerized band). 
If $t_1=t_2=t_3$ there is no induced spin density in
the transverse directions and therefore there is no quantization of the polar angle, since it is
ill-defined. These results are shown in Fig. \ref{trimer} where we show the topological bands and the
quantization of the polar angle when the spin-orbit term is present. The spin-orbit coupling induces
a transverse spin density and a quantization of the polar angle, even without the presence of an
external transverse magnetic field (a field along
$z$ splits the bands).
If $t_3\neq t_1=t_2$, we get in general non-trivial Berry phases, if
a spin-orbit coupling term is present. 
In Fig. \ref{t3lbd} we show the Berry phase (normalized by $2\pi$) 
as a function of $t_3i$ and $\lambda$, for zero transverse magnetic field but
with $B_z$ finite, showing the regimes where a non-trivial phase is found.
The spin-orbit coupling term 
acts as an effective non-local magnetic field (depends on momentum),
and tends to align the spin density. With $t_3=t_1=t_2$ the spin-orbit coupling does not lead to
non-trivial Berry phases, unless we add an external transverse magnetic field with the appropriate
direction. With $\lambda=0$ the addition of a field $B_x\neq 0$ (and $B_y$ satisfying the quantization
condition, Eq. (\ref{quantization})), also leads to a quantized
non-trivial Berry phase if $t_3>t_1,t_2$. With $t_3\neq t_1=t_2$, and with the spin-orbit term, we get in general
an alignment of the spin density. 

\begin{figure*}
\centering
\includegraphics[width=0.4\textwidth]{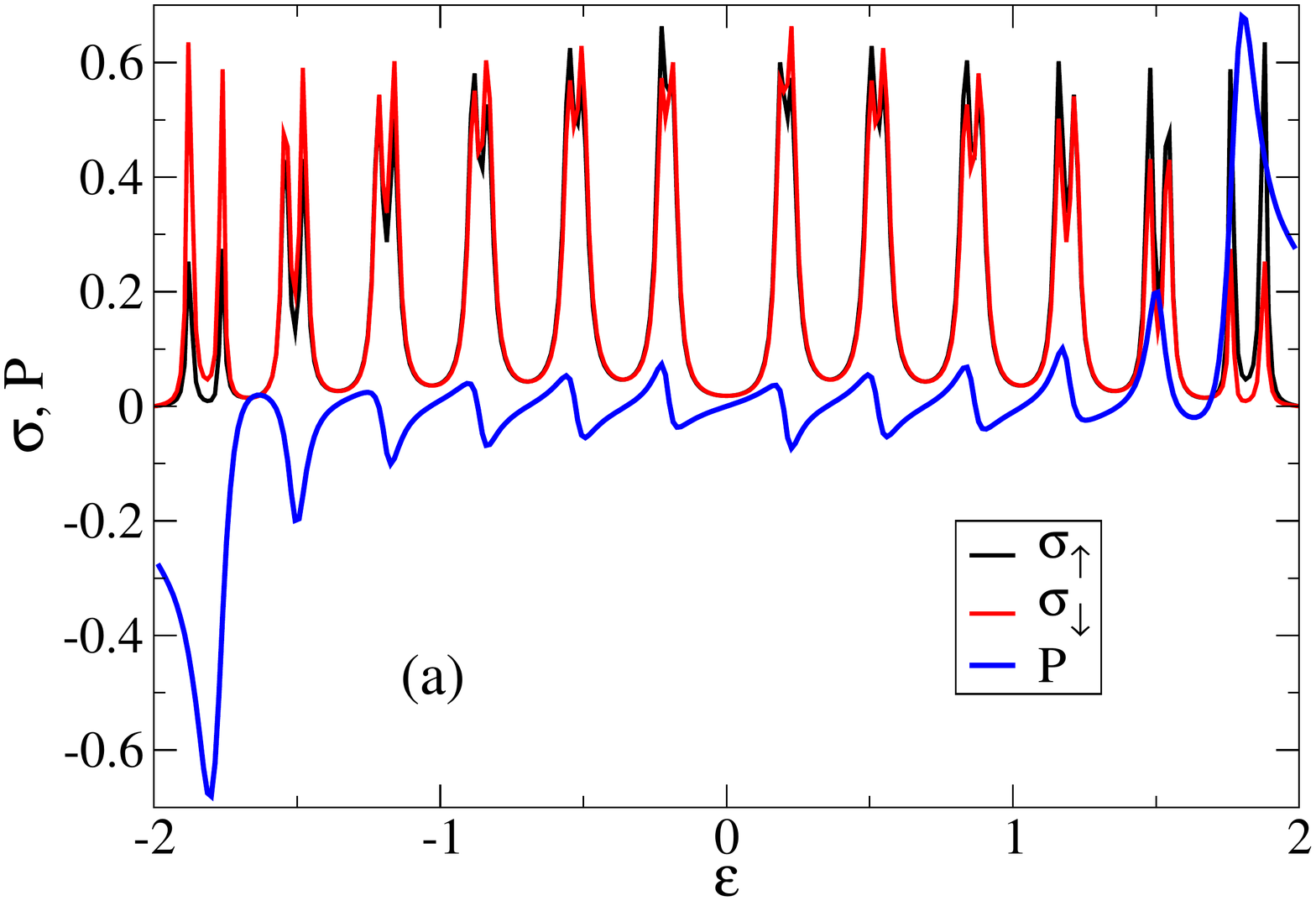}
\includegraphics[width=0.4\textwidth]{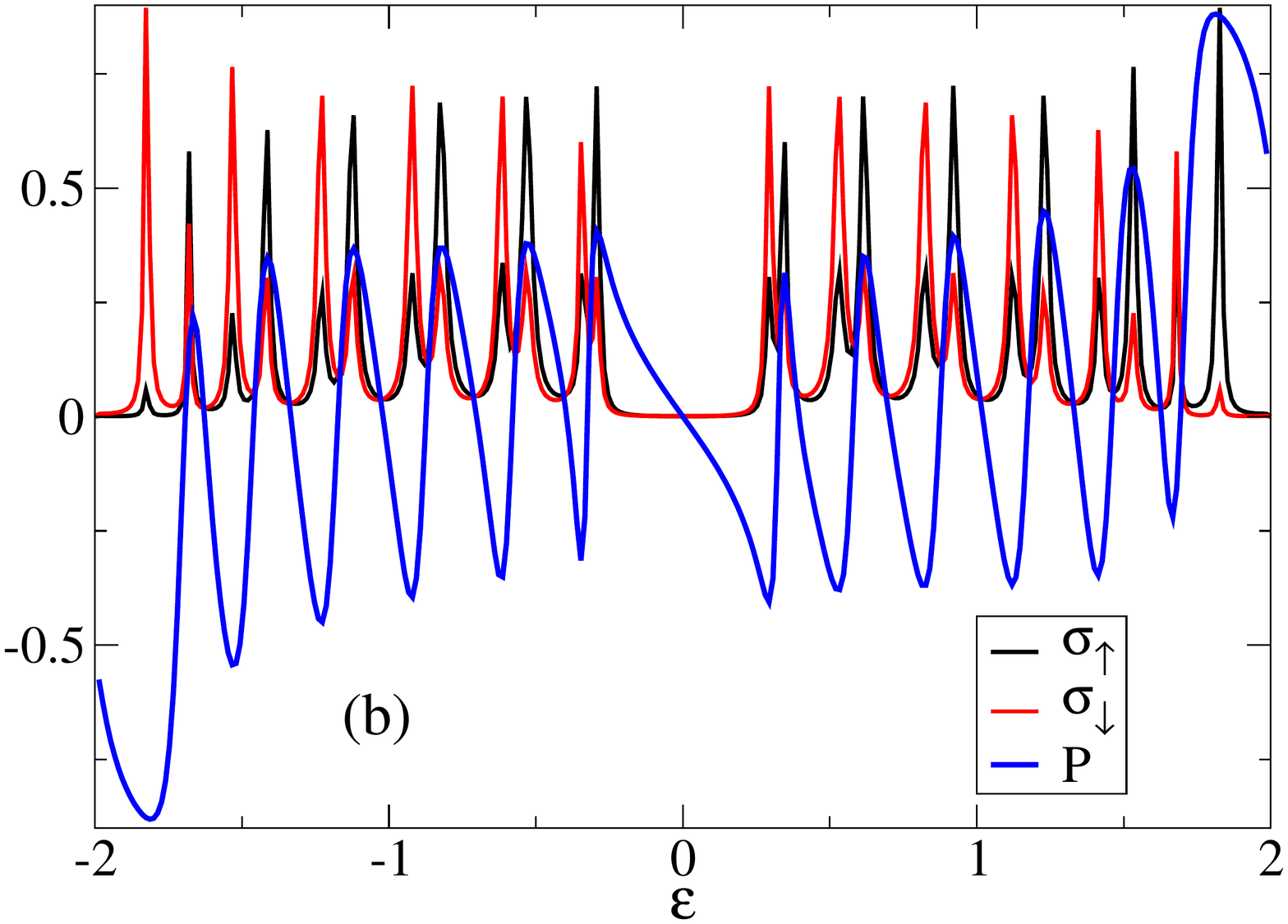}
\includegraphics[width=0.4\textwidth]{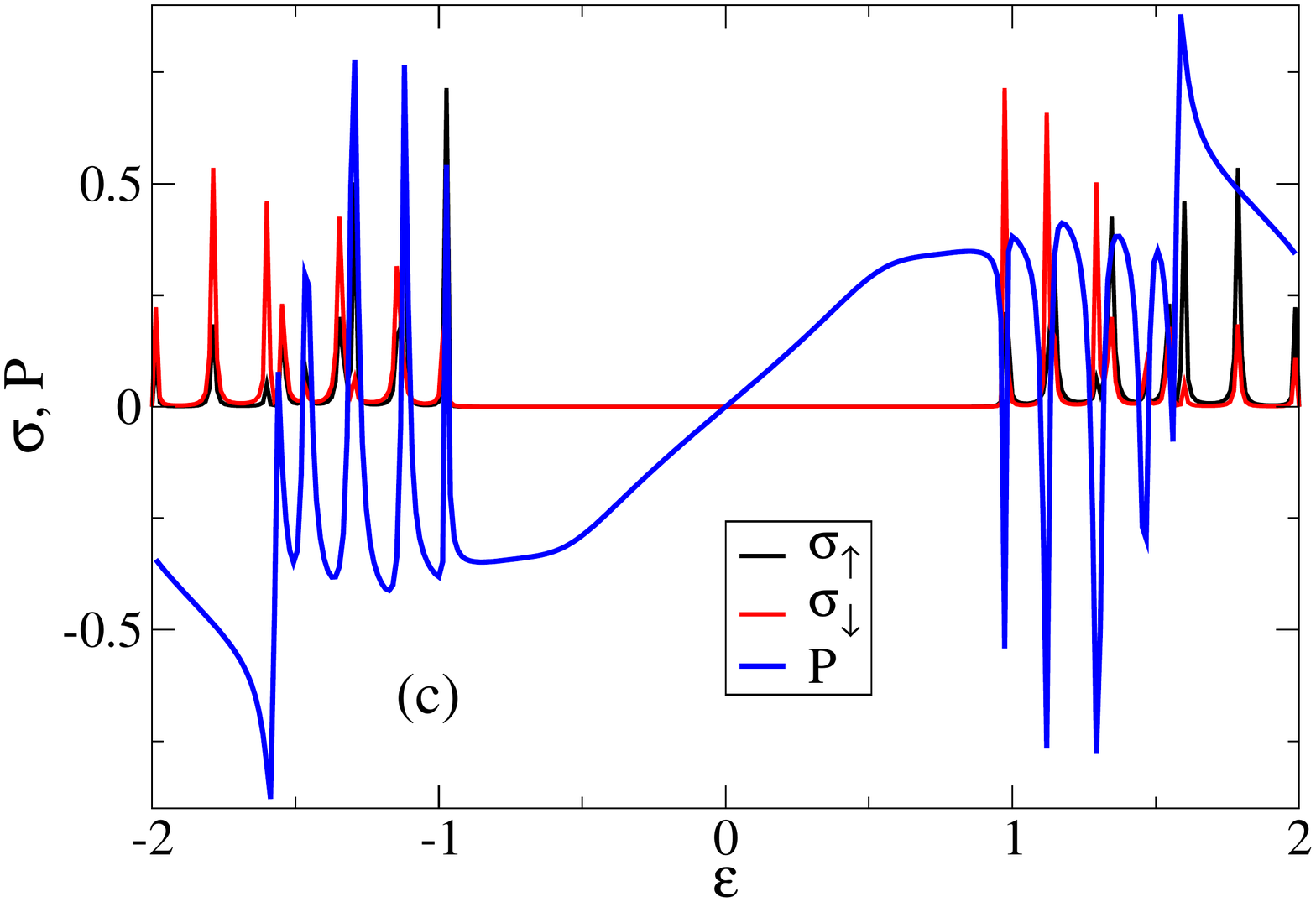}
\includegraphics[width=0.4\textwidth]{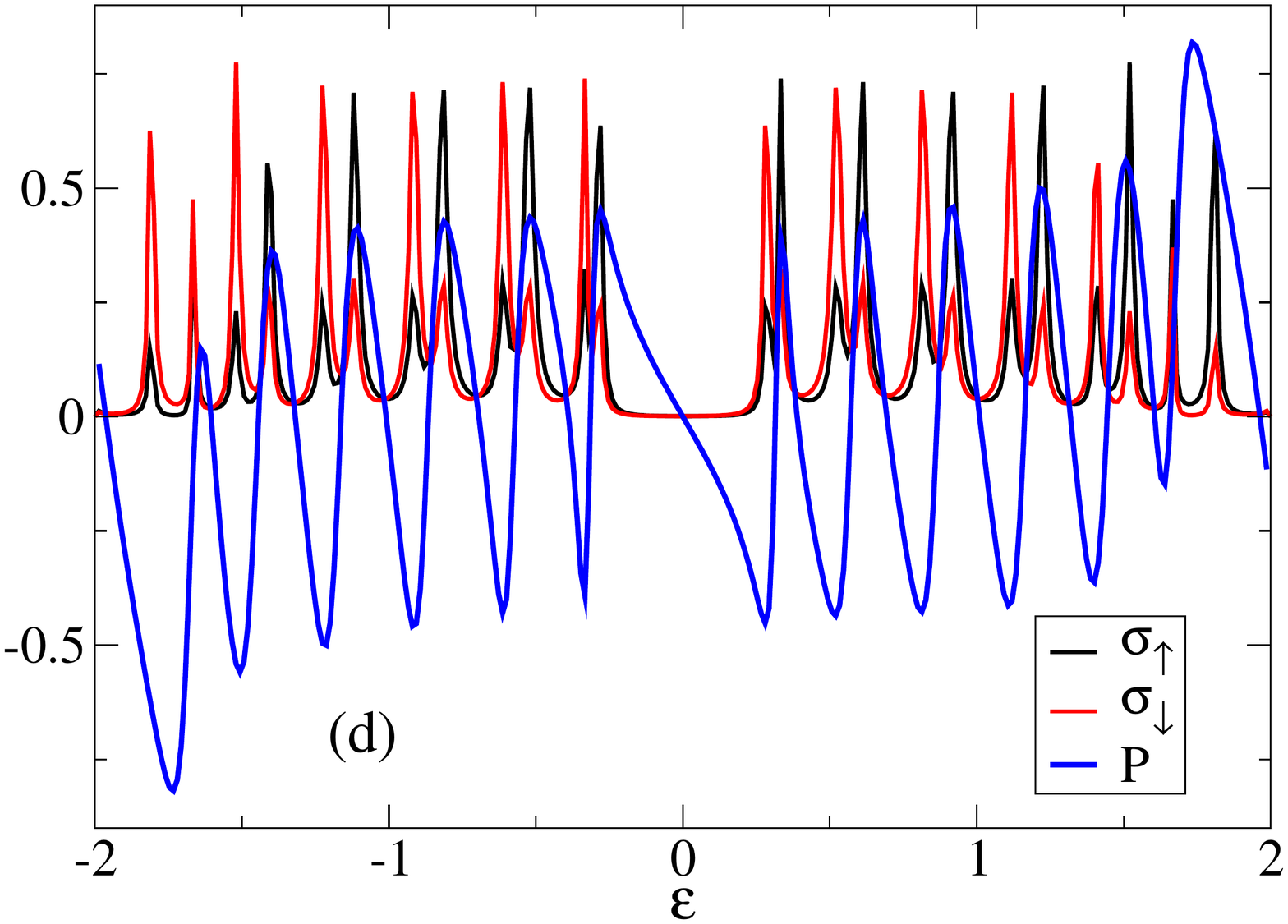}
\caption{\label{polarization}
Differential conductances for an incident electron with spin
$\uparrow$ and an incident electron with spin $\downarrow$ and spin
polarization, for $\gamma=4$, $10$ unit cells, $V=0.5,\lambda=2,B_z=0$ for
a) $B_x=0.5,B_y=0,\lambda_R=0$, b) $B_x=1,B_y=0, \lambda_R=0$, c) $B_x=1,B_y=0, \lambda_R=2$ and
d) $B_x=1,B_y=0.2, \lambda_R=0$.
}
\end{figure*}

We remark, however, that there is no spin polarization in spin transport 
if $\mathbf{B}=0$ since time reversal is preserved,
even though a spin density is created as a consequence of the spin-orbit coupling.
The spin density of an incoming spin up or spin down electron is rotated, but in such a way
that there is no difference between the two incoming electrons.

A related model with time reversal symmetry has been considered \cite{malard}
that includes Rashba, $\lambda_R$, and Dresselhaus, $\lambda_D$, spin-orbit couplings.
The Dresselhaus is uniform and the Rashba has a constant term plus a
modulated one, that can be argued to be in the class of a generalized Aubry-Andr\'e-Harper model.
There are several phases with winding number (chiral symmetry) and
gapless critical surfaces/lines and zero energy states. 
Inspired by these results we consider adding to the expression for $\chi(j)$ of our model Eq. \ref{Hamiltonian} a
constant $-E_y$ term, independent of the site, $j$, and a Rashba-like term that results from a constant $E_x$.
Adding for instance a spatially uniform Rashba-like term, with an amplitude $\lambda_R$,
leads to some consequences that will be explored in the next section.
If $\lambda$ and $\lambda_R$ are not simultaneously small, one finds zero energy states
if there is no external magnetic field. A term that may originate from a uniform electric field
along $-y$, may be considered with an amplitude $\lambda^{\prime}$. In the case of
zero external magnetic field and $\gamma=4$, the spectrum remains gapless around zero energy
if $\lambda_R=\lambda^{\prime}$ but there are two gaps between bands two and three and between
bands six and seven. If $\lambda_R \neq \lambda^{\prime}$ a gap opens around zero energy.
If $\lambda_R>\lambda^{\prime}$ there are edge states inside the three gaps with the appearance of
zero energy states. Applying an external magnetic field the zero energy states turn into
finite energy but are still well localized on the edges of a finite system, if the magnetic field is small enough. 

Terms of the type of a SSH model lead to zero energy states, in addition to the finite energy
edge states found in the helix problem.
The model with Rashba coupling \cite{malard} also leads to zero energy states (if $B=0$).
Adding $B_x$, the zero energy states in the gap acquire a small finite energy, but the states remain in
the gap that is centered around zero energy.
These states may contribute to either the low energy conductance or spin response when
the system is coupled to external leads. A significant effect is found if there is 
a small coupling to the leads and a not too small or too large system (see for instance, \cite{dong,balabanov}).
This will be considered next.

\section{Spin transport along a chiral chain}

\subsection{Scattering states}

Let us now determine the possible influence of the topological states on the
spin transport along the chain. We consider two leads that are attached to the
chiral chain. We may inject a spin up or spin down electron from the left lead
into the chiral chain and collect this scattering state on the right 
lead \cite{Landauer,houston}.
This implies, when time reversal symmetry is broken, a spin polarization on the
right lead.

We consider a one-dimensional problem that allows the propagation along the two spin channels.
The chiral region is of finite size. Due to the spin-orbit coupling and the transverse
external fields we must allow for spin flips. The scattering states may be reflected or
transmitted along the chiral region, keeping the same spin component or reversing the
spin component. In general we have a $2 \times 2$ matrix structure, respecting to the two
spin projections. The scattering states propagate along the chiral region. We choose the
left and right leads as simple conductors, using a tight-binding model, with bandwidth
large enough to go over the energy range of the chiral region. The hopping from the leads
to the chiral region is parametrized by $V$, in units of $w$.

At each lattice site, $n$, we have in general a spinor, $\psi_n$, respecting to the two spin projections leading
to an equation of the type
\be
H_n \psi_n + T_{n,n+1} \psi_{n+1} + T_{n,n-1} \psi_{n-1} = \epsilon \psi_n
\ee
Here, $\epsilon$ is the energy of the electron coming from the left lead. We may rewrite this equation as
\be
\psi_{n+1} = T_{n,n+1}^{-1} \left( \epsilon -H_n \right) \psi_n - T_{n,n+1}^{-1} T_{n,n-1} \psi_{n-1}
\ee
which leads to a transfer matrix, $M_n$, defined as
\bea
\label{mn} 
\left(\begin{array}{c}
\psi_{n+1} \\
\psi_n
\end{array}\right)
&=& M_n 
\left(\begin{array}{c}
\psi_{n} \\
\psi_{n-1}
\end{array}\right)
\nonumber \\
M_n &=&
\left(\begin{array}{cc}
T_{n,n+1}^{-1} \left( \epsilon-H_n \right) & -T_{n,n+1}^{-1} T_{n,n-1} \\
1 & 0
\end{array}\right)
\nonumber \\
\eea
In units of the electric field ($E=1$) we find that the determinant of $M_n$ is unity.

Let us consider the combined system of the two leads and the chiral region and take the
size of the combined system as $N$. We may consider a point on the left lead, at position
$N_L$ sufficiently far from the chiral region and a point on the right lead, $N_R$, also sufficiently
far from the chiral region. We may write that
\begin{equation}
\label{mt} 
\left(\begin{array}{c}
\psi_{N_R+1} \\
\psi_{N_R}
\end{array}\right)
= M_T
\left(\begin{array}{c}
\psi_{N_L} \\
\psi_{N_L-1}
\end{array}\right)
\end{equation}
where
\be
M_T = \prod_{n=N_L}^{N_R} M_n
\ee
Considering either an electron coming from the left lead with spin up or spin down, we
may obtain the differential conductance and the spin polarization in a standard way,
as briefly reviewed in Appendix B.

\begin{figure}
\centering
\includegraphics[width=0.4\textwidth]{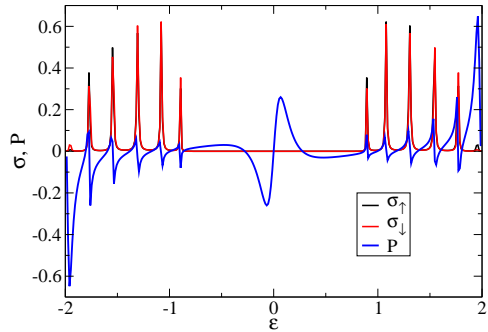}
\caption{\label{edgec}
Differential conductances and spin polarization for $\gamma=4$, $10$ unit cells,
$\lambda=2,\lambda_R=2,B_x=0.1$. In this case the Berry phase is not quantized but
due to the addition of a constant Rashba spin-orbit coupling, there is a low-energy
state that is shifted from zero energy due to the presence of a finite $B_x$. The effect
of the edge state in the gap is clearly seen in the spin polarization.
}
\end{figure}

\begin{figure}
\centering
\includegraphics[width=0.4\textwidth]{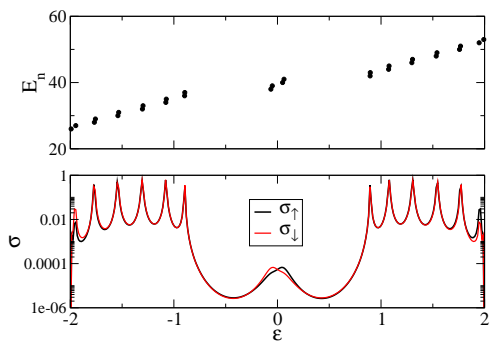}
\caption{\label{peaks}
Location of peaks of differential conductance and of eigenenergies of finite
system of $10$ unit cells, $E_n$, uncoupled to the leads. 
The parameters used are $\lambda=2, \lambda_R=0, B_x=0.5, V=0.1$. 
}
\end{figure}

\begin{figure*}
\centering
\includegraphics[width=0.4\textwidth]{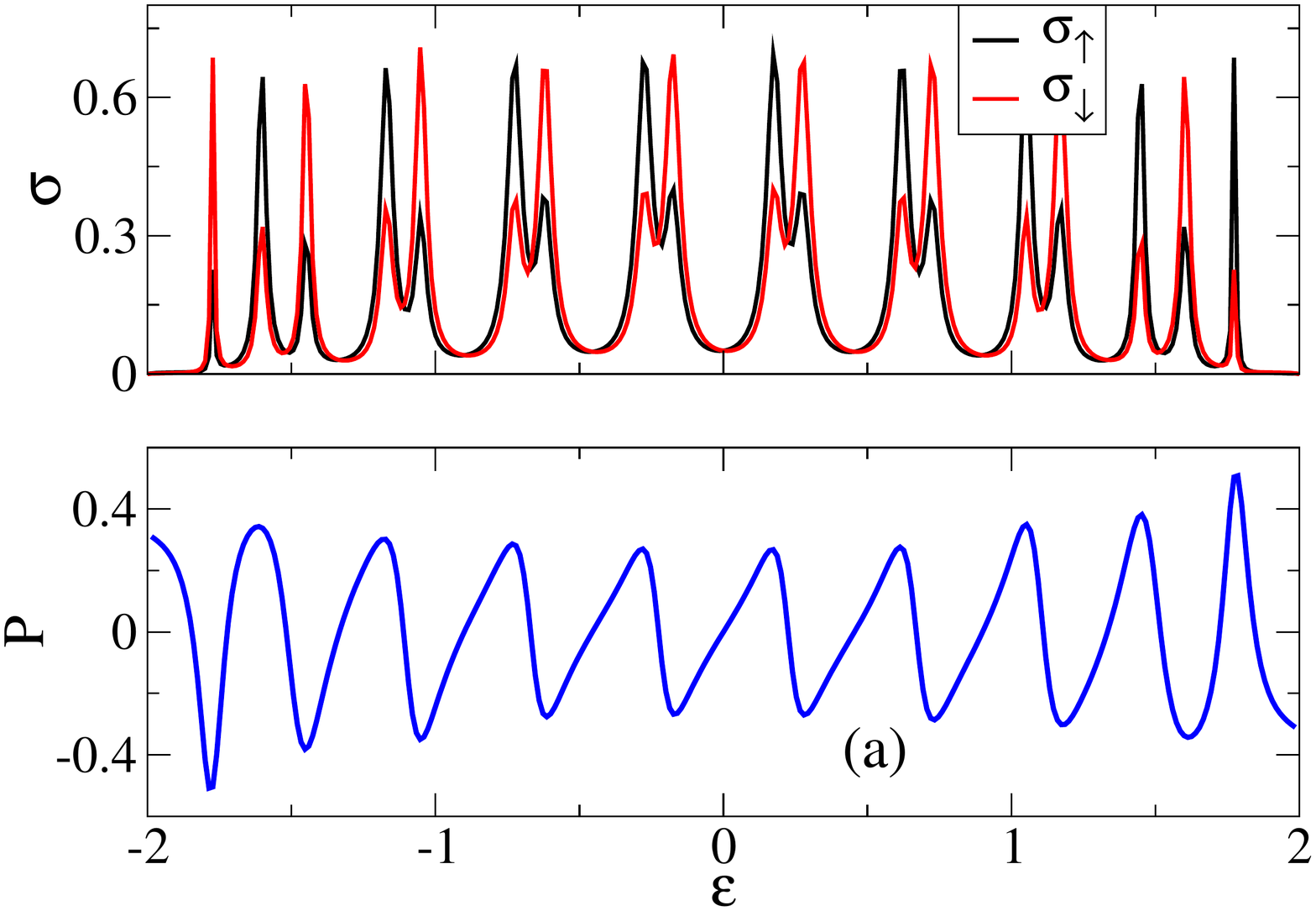}
\includegraphics[width=0.4\textwidth]{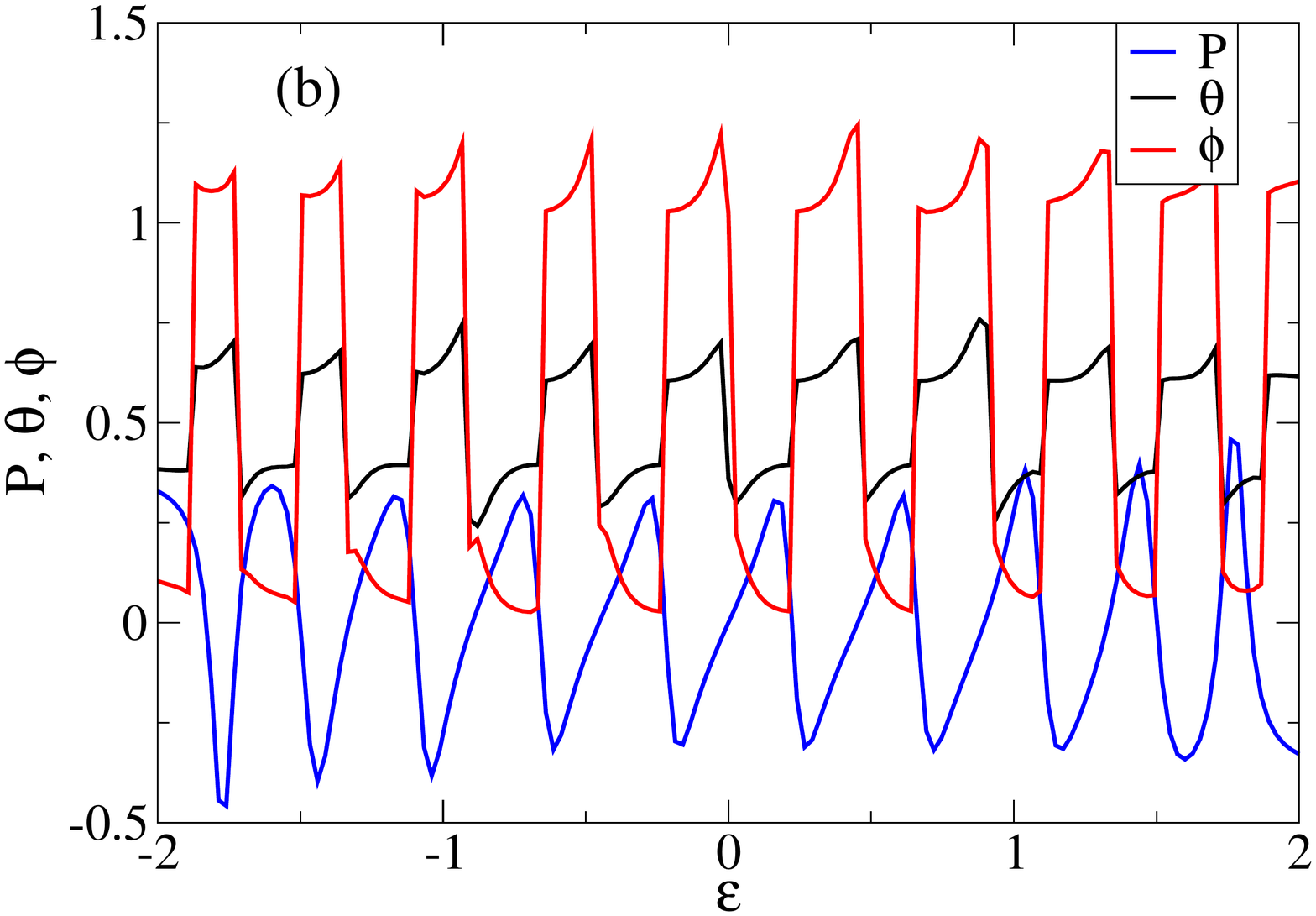}
\includegraphics[width=0.4\textwidth]{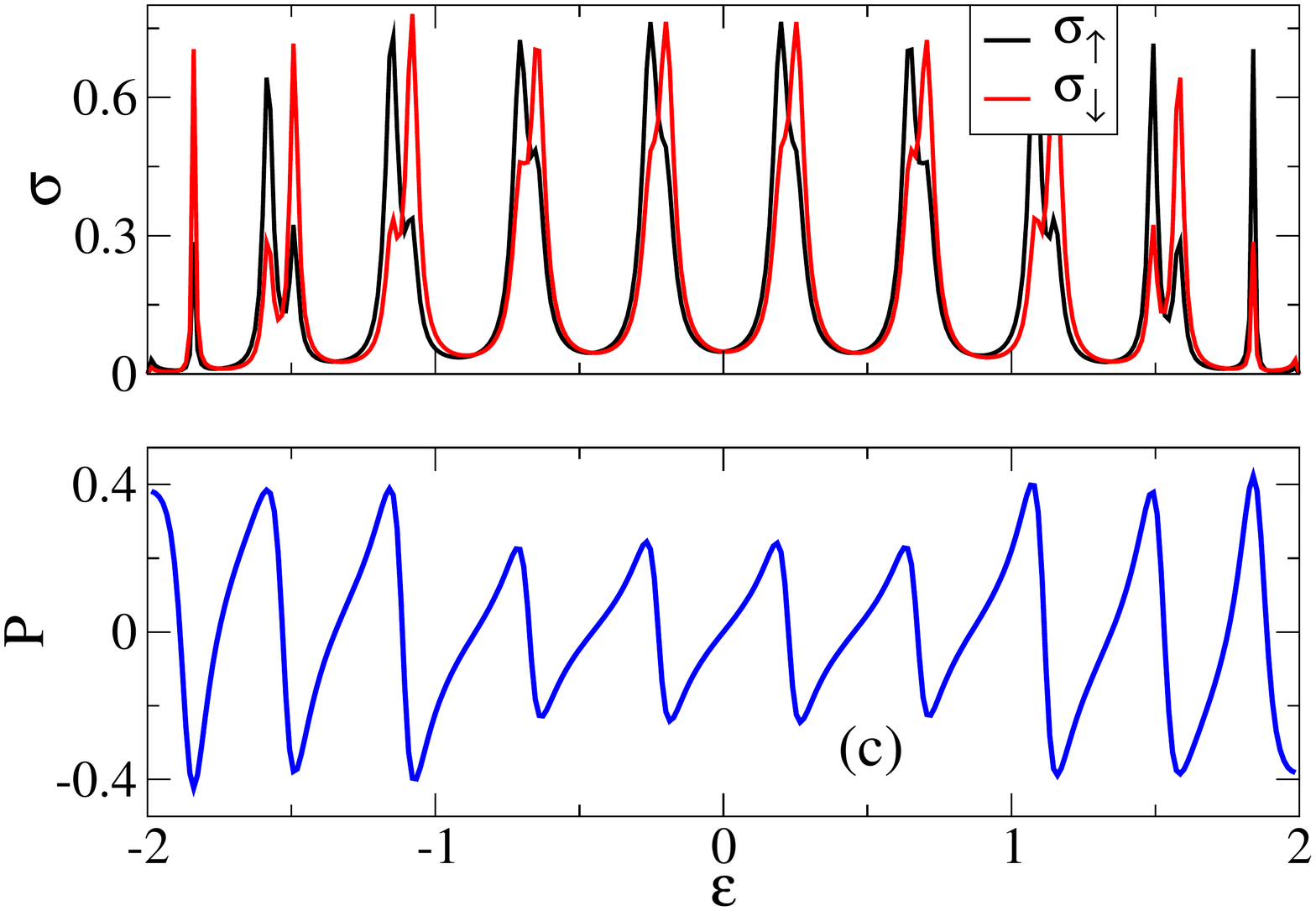}
\includegraphics[width=0.4\textwidth]{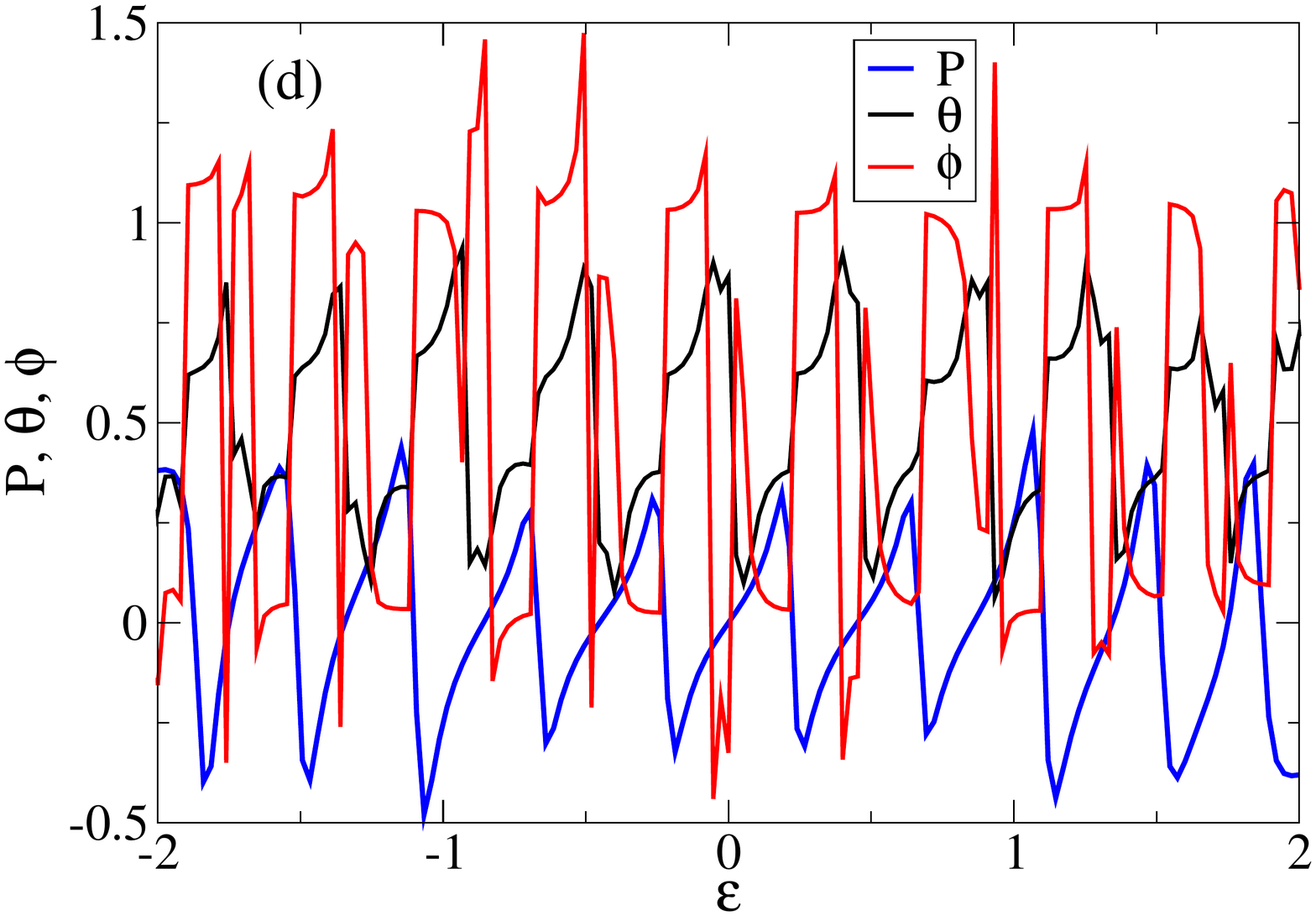}
\caption{\label{smoothing}
a) and c) Differential conductances and spin polarization and b) and d)
angles $\theta$ and $\phi$ of the spin density at the right edge
of the chiral chain for topological, (a) and b), and non-topological, c) and d), cases. 
The parameters are $\gamma=3,\lambda=2,B_x=0.5,V=0.5$ and $10$ unit cells for
$B_y$ given by condition Eq. (\ref{quantization}) or $B_y=0$.
}
\end{figure*}

The charge differential conductances, for incident electrons with spin components
$\uparrow,\downarrow$ are obtained as
\bea
\sigma_{\uparrow} &=& 1- |r_{\uparrow}|^2 - |\bar{r}_{\uparrow}|^2 = 
|t_{\uparrow}|^2 + |\bar{t}_{\uparrow}|^2
\nonumber \\
\sigma_{\downarrow} &=& 1- |r_{\downarrow}|^2 - |\bar{r}_{\downarrow}|^2 = 
|t_{\downarrow}|^2+ |\bar{t}_{\downarrow}|^2
\eea
The spin polarization is defined as
\be
P=\frac{\left( |t_{\uparrow}|^2+|\bar{t}_{\uparrow}|^2 \right) -
\left( |t_{\downarrow}|^2+|\bar{t}_{\downarrow}|^2 \right)}{
\left( |t_{\uparrow}|^2+|\bar{t}_{\uparrow}|^2 \right) +
\left( |t_{\downarrow}|^2+|\bar{t}_{\downarrow}|^2 \right)}
\label{definition}
\ee
The reflection (transmission) coefficients $r_{\sigma}$ ($t_{\sigma}$) of an incident
electron with spin $\sigma$ with no spin flip, and the coefficients with
spin flip $\bar{r}_{\sigma}$ ($\bar{t}_{\sigma}$) are defined in Appendix B.

\subsection{Results}

Applying a magnetic field along the chain direction naturally leads to a spin polarization
along that axis. In the following we consider that $B_z=0$ and only consider the action of 
a transverse magnetic field.
In Fig. \ref{polarization} we present results for the spin polarization 
and the charge diferential conductances, $\sigma_{\uparrow},\sigma_{\downarrow}$ for a
chiral chain with $10$ cells with a unit basis of four atoms, $\gamma=4$. As discussed in
Appendix B, the effect of any edge states is lost if the chain is large enough, since they
decay exponentially towards the interior of the chain. Also, the size of the hopping from the
leads to the chiral chain has to be chosen appropriately. We compare the situations when
the Berry phase is quantized and when it is trivial. The trivial cases are obtained either
adding a Rashba term or considering $B_y \neq 0$. In a) and b) two different values of
$B_x$ are considered. As $B_x$ increases from $B_x=0.5$ to $B_x=1$, the spin polarization
increases in magnitude and the gap centered around zero energy also increases.
Adding a constant Rashba term in c) increases the gap and comparing b) with d) we see that adding a small
$B_y=0.2$ does not affect significantly the spin polarization. Note that the various peaks
of the spin polarization correlate well with the peaks in the conductances.  

\begin{figure}
\centering
\includegraphics[width=0.4\textwidth]{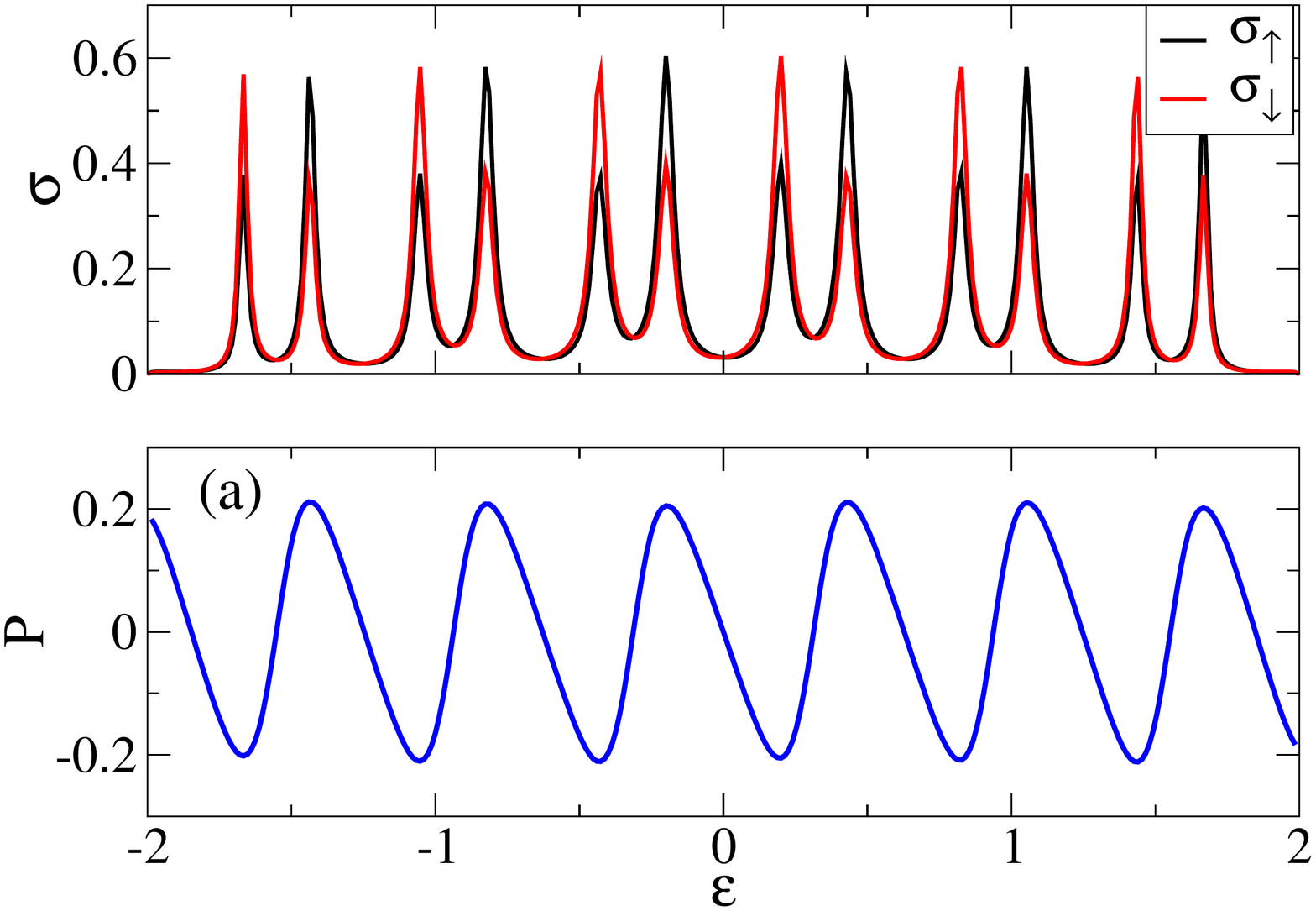}
\includegraphics[width=0.4\textwidth]{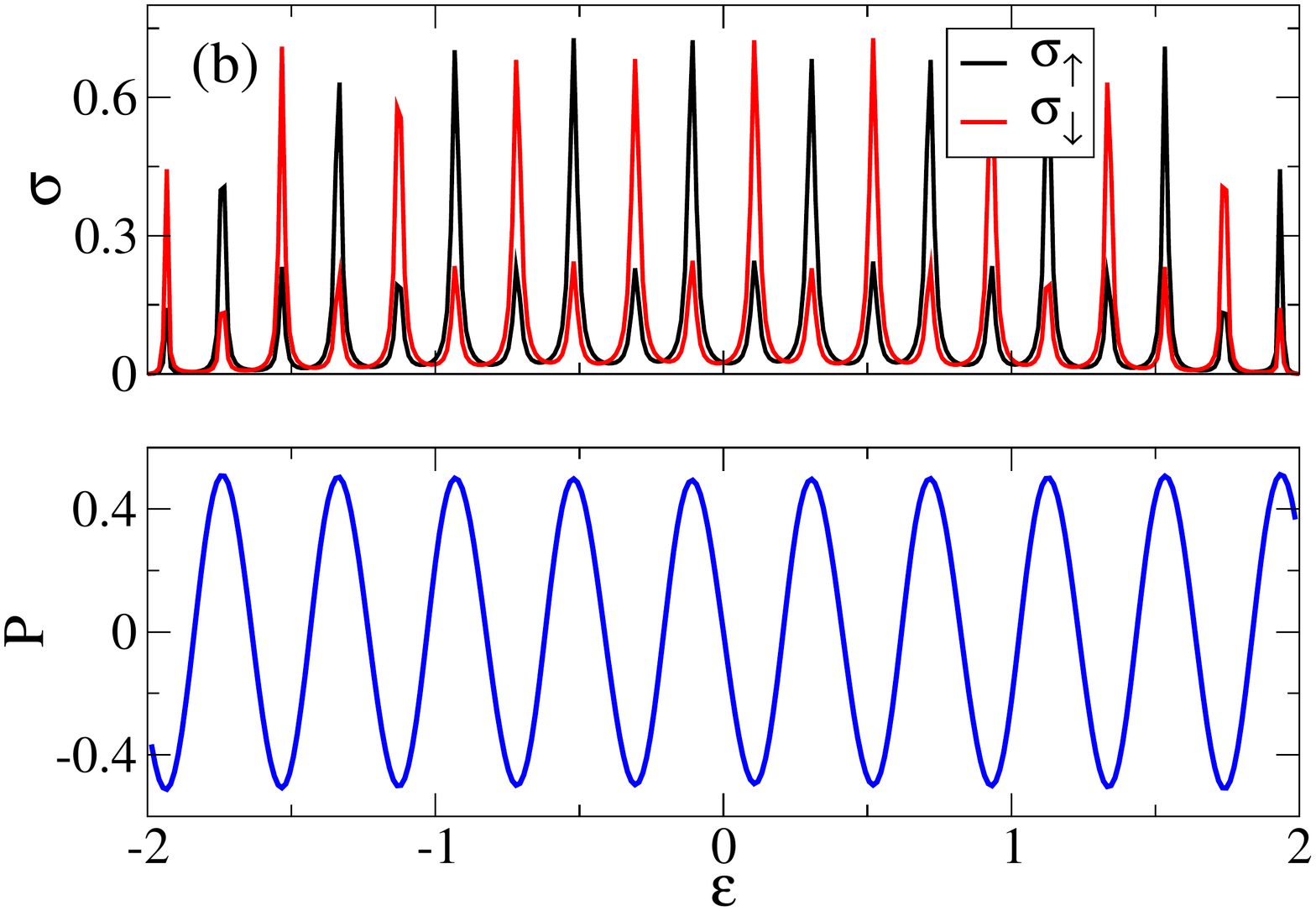}
\caption{\label{periodics}
Differential conductances and spin polarization for two sets of parameters
that yield an almost periodic spin polarization as the energy changes.
The sets of values are for $\gamma=4,V=0.5,B_x=B_y=0.1$ and 
a) $10$ cells, $\lambda=2, \lambda_R=3, \lambda^{\prime}=3$ and
b) $20$ cells, $\lambda=3, \lambda_R=4,\lambda^{\prime}=4$. 
}
\end{figure}

In Fig. \ref{polarization}(c) a finite Rashba coupling, $\lambda_R=2$, is added. 
This leads, in the finite chiral chain
uncoupled to the leads, to the presence of zero energy edge states if $B_x=0$, that have a
finite energy since $B_x \neq 0$. Since $B_x$ is large in this case, the state energy is close to the gap
and its effect is not seen.
In Fig. \ref{edgec} we consider a smaller value of $B_x=0.1$ and its presence is clearly seen
in the spin polarization at a small energy inside the gap, of the order of $5 \times 10^{-2}$. 
The contribution of the edge
states to the differential conductances is however small, and we need to lower the coupling to the
leads, $V$, to see their effect, as shown in Appendix B. In addition to the oscillation in
the spin polarization near zero energy, the peaks of the spin polarization correlate
well with the peaks of the conductances, as in the previous figure.
In Fig. \ref{peaks} it is shown that the peaks of the differential conductances are located
at the eigenenergies of the finite size chiral chain, uncoupled to the leads, with open boundary conditions.
Recall that the energy $\epsilon$ considered is not an eigenenergy of the junction and indeed
the states are scattering states. Still the resonance with the eigenstates of the uncoupled
chain is clearly seen.

The results in Fig. \ref{polarization} show that the spin polarization is not significantly
affected by the Berry phase quantization,
even though its effect is felt in the polar angle of the spin density in the chiral chain, 
both in momentum space and in real space. However, let us analyze the situation in some detail.
In Fig. \ref{smoothing} we consider a chain with $\gamma=3$. The quantized Berry phases are found
imposing the condition Eq. (\ref{quantization}). A simple trivial case is obtained taking $B_y=0$.
The top panels of Fig. \ref{smoothing} refer to a quantized Berry phase and the lower panels
to the trivial case. In Fig. \ref{smoothing}(a),(c) we show the differential conductances and
the spin polarization and in Fig. \ref{smoothing}(b),(d) we show the angles $\theta$ and $\phi$ of
the spin density at the right end of the chiral chain. The energies for which $\theta<\pi/2$
($\theta>\pi/2$) correspond to the regimes where the spin polarization is positive (negative),
as expected. Comparing the structure of the conductance peaks we see a similar structure for
the quantized and trivial cases (and similar behaviors for the spin polarization) even though it may
be argued that in the quantized case the behavior is somewhat more regular.
This is emphasized considering the energy dependence of the angles $\theta,\phi$ where
a sharper behavior is found in the quantized (topological) case.

A spin polarization that is nearly periodic in the incident energy,
may be obtained 
symmetrizing the spin-orbit coupling (considering equal Rashba, $\lambda_R$,
and $\lambda^{\prime}$ constant terms, which leads to a gapless spectrum around
zero energy) and symmetrizing the transverse external
magnetic field (taking $B_x=B_y\neq 0$). In Fig. \ref{periodics} we consider
two examples that show a rather regular sequence of energy peaks in the differential conductances
(and their energy spacings), that lead to a regular set of oscillations and amplitudes
of the energy dependent spin polarization. We consider $\gamma=4,V=0.5$ and take
$B_x=B_y=0.1$. Considering a chain with $10$ unit cells the example shown considers
$\lambda=2,\lambda_R=\lambda^{\prime}=3$ and considering $20$ unit cells we take
$\lambda=3, \lambda_R=\lambda^{\prime}=4$. We see that increasing the system size the polarization is 
very nearly periodic.

\section{Conclusions}

The spinful electronic states of the multiband chiral chain, with a unit cell of size $\gamma$, 
are the result 
of the electric field/electric polarization felt by a moving electron
that is seen, in the reference frame of the electron, as a modulated spin-orbit coupling.
The spectrum is in general gapless.

One way to split the levels is to apply an external magnetic field. Applying a field along
$z$ partially lifts the degeneracy and adding a field in the plane transverse to the
chain axis, lifts the degeneracies of all bands, giving origin to a gapped system.
Turning off $B_z$, it is also possible to obtain a gapped system in which the bands are grouped
in pairs. One advantage of applying the magnetic field is that it breaks time reversal invariance
and allows the possibility of spin transport. Another possible outcome is the possibility
of non-trivial topological states as a result of gap openings.

Under appropriate circumstances the gapped system has bands with quantized Berry phases of either
individual bands or, in the case of $B_z=0$, of pairs of bands. This suggests a topological nature
of the bands, in these regimes. The topological character is also revealed by the quantization
of the polar angle of the spin density. If the Berry phase is not quantized the polar angle
is also not quantized. This is also found diagonalizing the problem in real space, with open
boundary conditions, since the polar angle of the spin density averaged over a (bulk) eigenstate
of the finite chain is also quantized.
We have argued that the quantization of the Berry phase in the case of $\gamma=4$ may be associated with
a (hidden) symmetry that is the product of an inversion symmetry and a chiral symmetry, even though
these symmetries are absent separately. 
Edge states are found, displaying very
low participation ratio and well localized wave functions on the edges of a finite chain.
However, these edge states have finite energies and not zero energy.
Moving away from the quantized conditions for the Berry phase and the spin polar angle,
we also find edge states. A difference is found that in the quantized Berry phase regime the edge states
appear in pairs that are energy degenerate, while in the trivial regime the states are no longer
degenerate  and either appear as a single state (as for $\gamma=3$) or as consecutive
energy states with a finite energy separation (as for $\gamma=4$), at least for the finite sizes
considered. Another relevant difference is that if the system size is large enough, the polar angles of the
spin density
of the edge states are different from each other (between the two degenerate states) and are
not quantized. 

The gapless chiral model with no external magnetic field may be gapped adding a homogeneous Rashba-like term,
revealing the existence of topology and localized zero energy edge states. Adding the external magnetic field
to the chiral model with no Rashba-like term,
the edge states remain but acquire a finite energy and under appropriate conditions the Berry phase is quantized.
Combining the two effects of external magnetic field and Rashba-like term, the model is not
topological in the conventional sense, the Berry phase is not quantized, but the edge states remain,
both at small and finite energies, and time-reversal symmetry
is broken. The breaking of time-reversal symmetry allows a spin polarization in the transport along
the chiral chain, and the vicinity to a topological phase leads to the existence of the low energy
states in the gap.

We have also considered models where the lattice may be distorted that lead to gap
openings in appropriate conditions. For instance, a unit cell with
two sites, $\gamma=2$, which may be dimerized, or a unit cell with $\gamma=3$, which may be trimerized in some
way. The first case is similar to a SSH model (including spin). The model by itself has a topological
regime, with zero energy edge states, quantized winding number and quantized Berry phase.
In the case of $\gamma=3$ we showed explicitly that if in addition to the trimerization we add the
spin-orbit coupling, even in the absence of an external transverse magnetic field (a field along
$z$ is useful to split the bands), a transverse spin density is induced and a quantization of
the spin polar angle is found. 
We also found a quantized Berry phase for a model of electrons with no spin,
but with a non-trivial hopping structure, due to the presence of more that one orbital state and considering
both nearest neighbor and next-nearest neighbor terms (with a complex part), as shown in Appendix A.

In the case the Berry phase is quantized, the energy dependence of the spin polarization
has a better defined structure in comparison to a general case with no Berry phase quantization.
Adding a Rashba-like homogeneous spin-orbit term, the contribution 
of the near zero energy edge states leads to an increase of the 
low-energy spin polarization. 
Also, we found a nearly energy periodic structure for the spin polarization
considering symmetric homogeneous couplings and a symmetric transverse
transverse magnetic field.

As a final note the helix may also be seen as a ladder where one chain has spin up and the
other spin down and it may be related to other models.
Alternating spin up and spin down the helix is related to a Creutz ladder \cite{creutz,platero}:
$\gamma=4$ a staggered Creutz ladder with fluxes $0$ and $\pi$.
For $\gamma=3$ trimerized ladder with fluxes $0, 2\pi/3, 2\pi/3, 0, -2\pi/3, -2\pi/3, 0$.

\acknowledgements
We thank discussions with Pedro Ribeiro, Miguel Ara\'ujo, Henrik Johannesson, Mariana Malard, Bruno Mera, Jamal
Berakdar and Vitalii Dugaev.
We also acknowledge partial support from FCT through the Grant UID/CTM/04540/2019.

\begin{figure}
\centering
\includegraphics[width=0.4\textwidth]{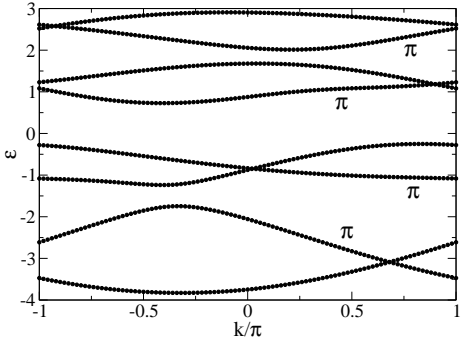}
\caption{\label{orbital}
Energy bands and Berry phases of Hamiltonian Eq. (\ref{Horbital}) 
for $t_1=1, t_2=0.5, t_{\perp}=1, \Delta(1)=-1, \Delta(2)=1$.
}
\end{figure}

\appendix

\section{Orbital coupling}

Another possibility that has been considered is to ignore the spin contribution and to
look for an orbital origin \cite{liu}. In this context, at each molecule in the helix one
considers different states associated with the orbital degrees of freedom. The chiral
nature of the molecule is implemented introducing a twist (phase) in the hoppings
between different orbitals of neighboring molecules. One may understand such a construction
considering now that the electron actually moves along the helix and therefore has
momentum components along the $xy$ plane. Under the action of the electric field the
effective magnetic field has now a $z$-component (along the axis of the helix)
that may be seen as a flux that gives origin to an orbital effect (due to the
corresponding vector potential) that may be incorporated as a Peierls factor in
the effective hoppings between the molecules. To simplify one may just consider that at
each molecule two states of orbital nature contribute.

\begin{figure*}
\centering
\includegraphics[width=0.32\textwidth]{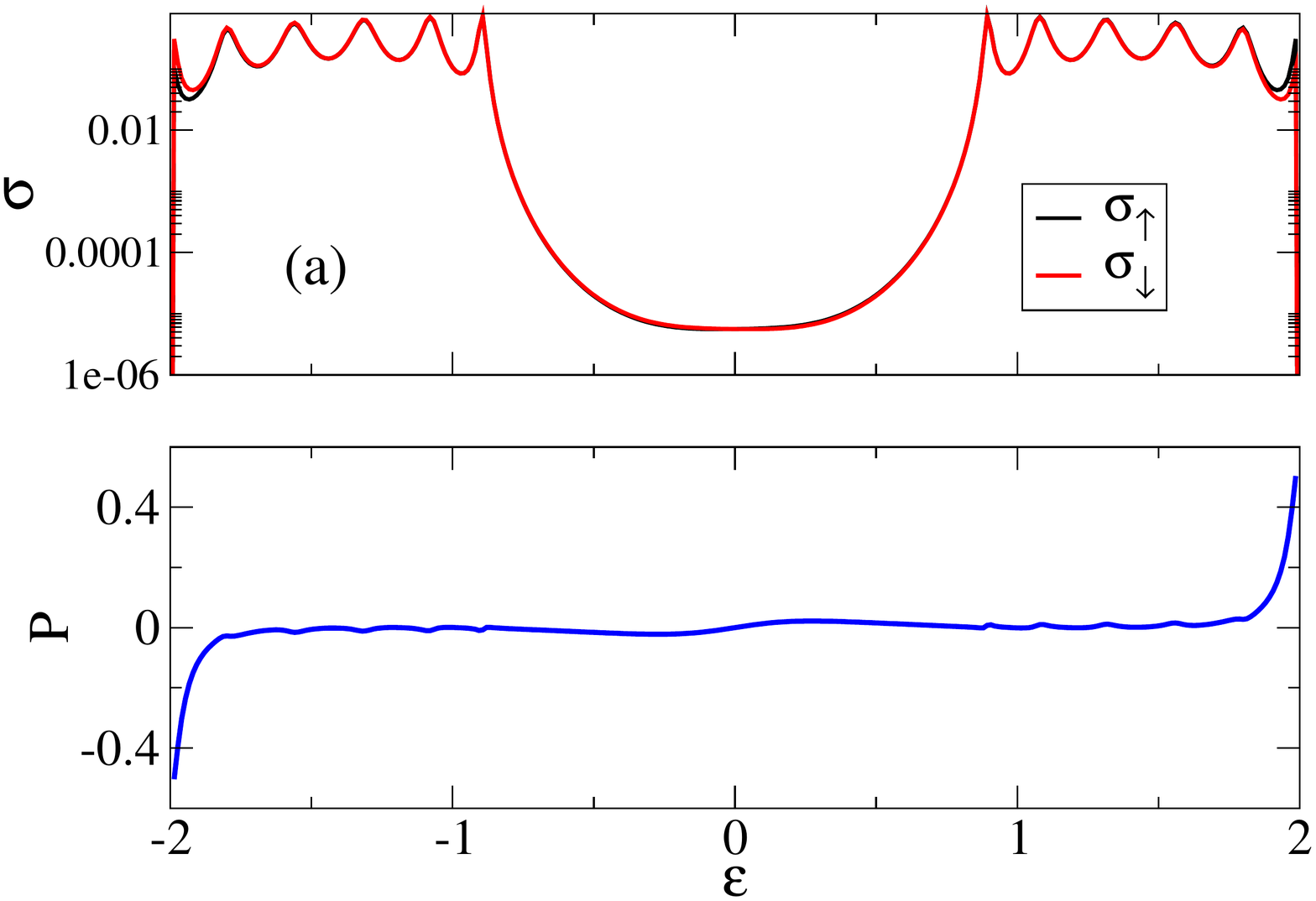}
\includegraphics[width=0.32\textwidth]{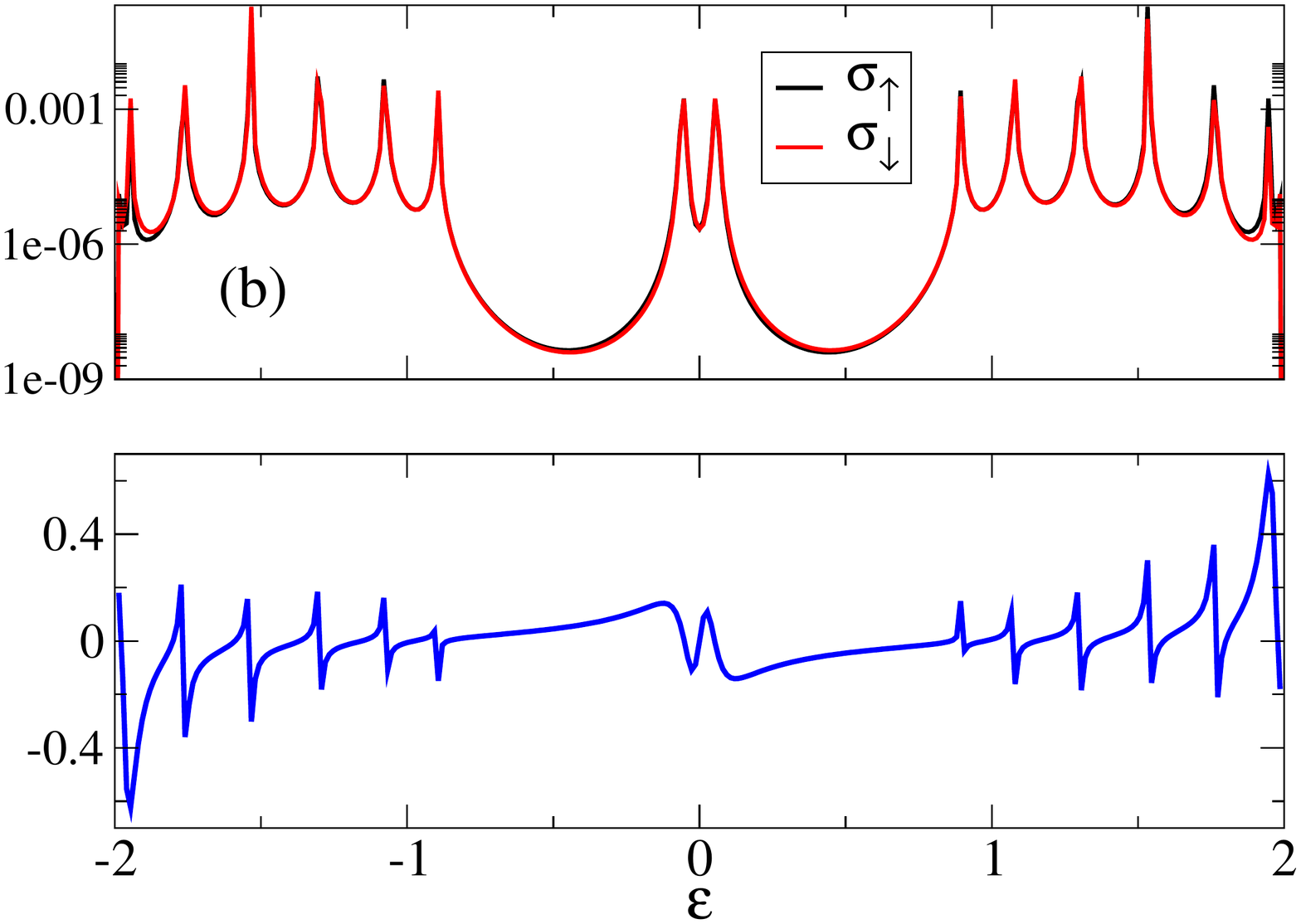}
\includegraphics[width=0.32\textwidth]{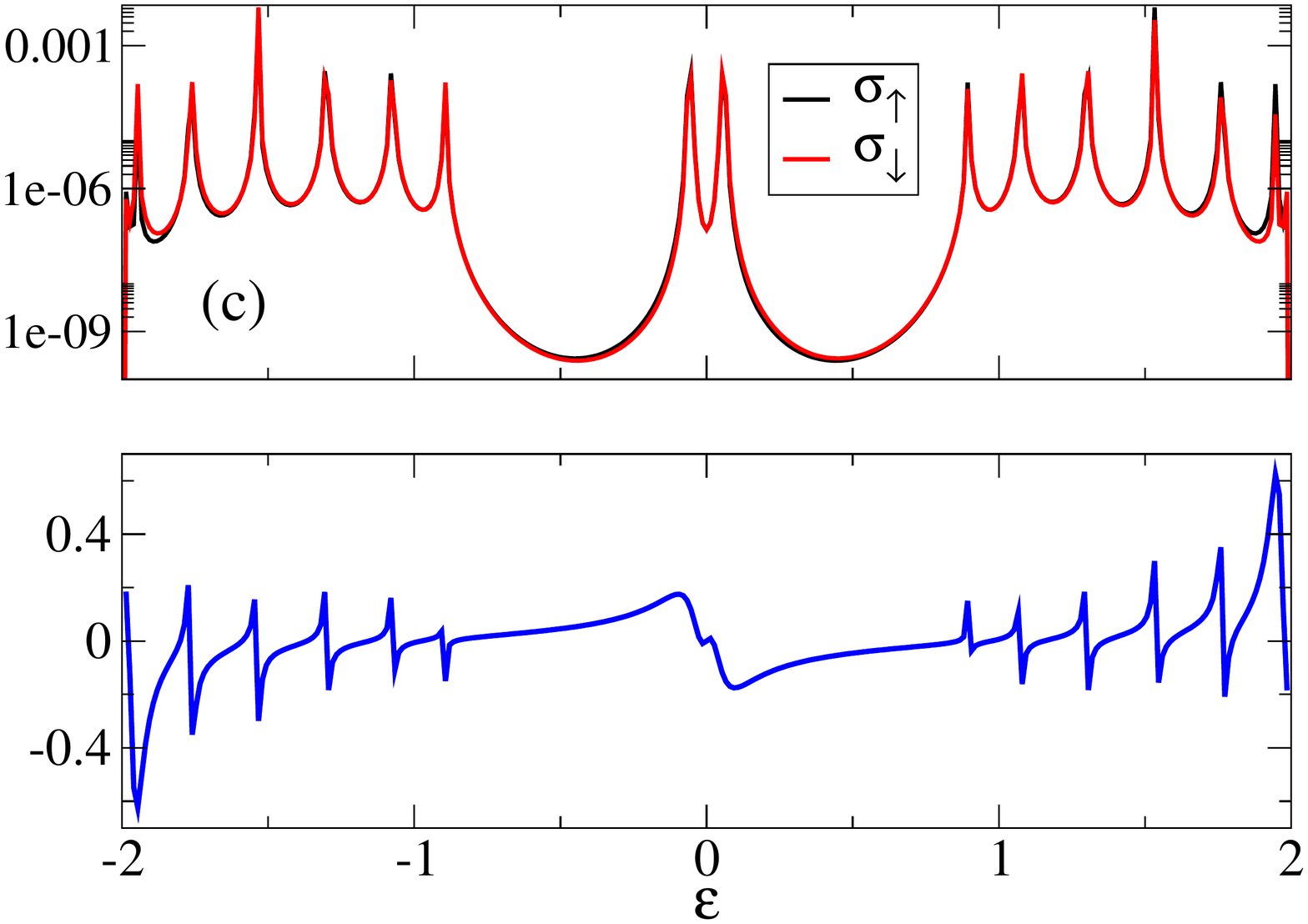}
\includegraphics[width=0.32\textwidth]{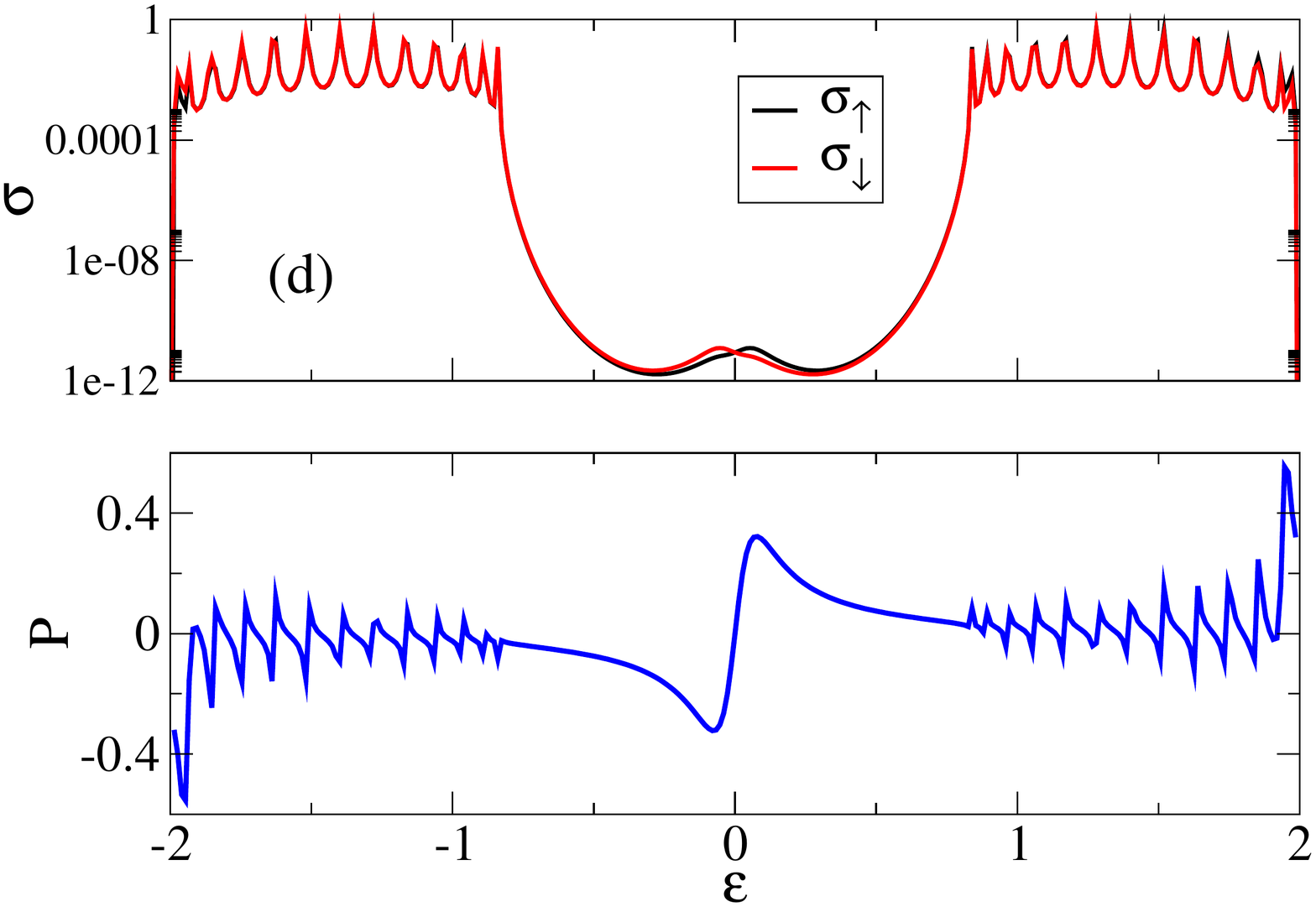}
\includegraphics[width=0.32\textwidth]{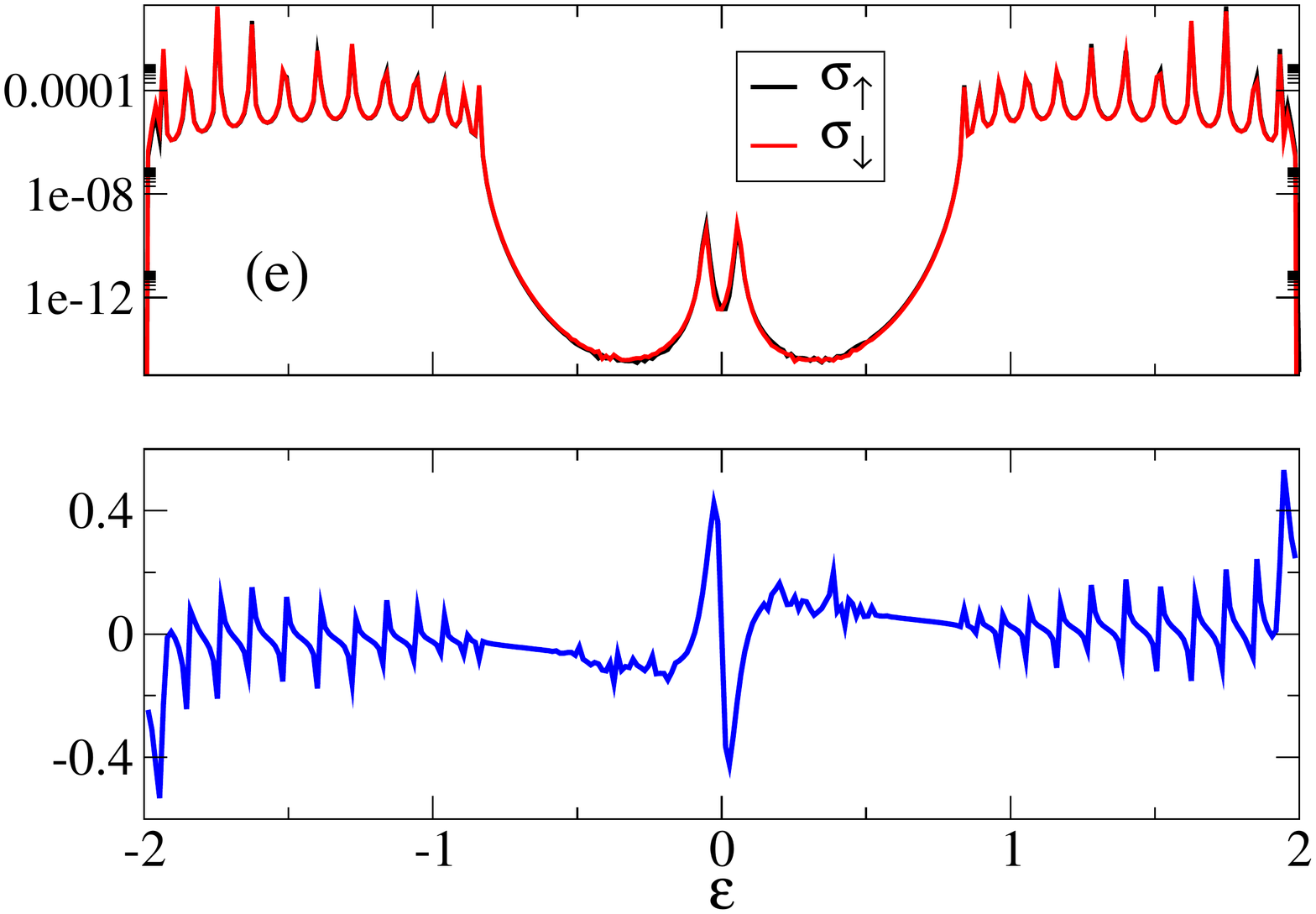}
\includegraphics[width=0.32\textwidth]{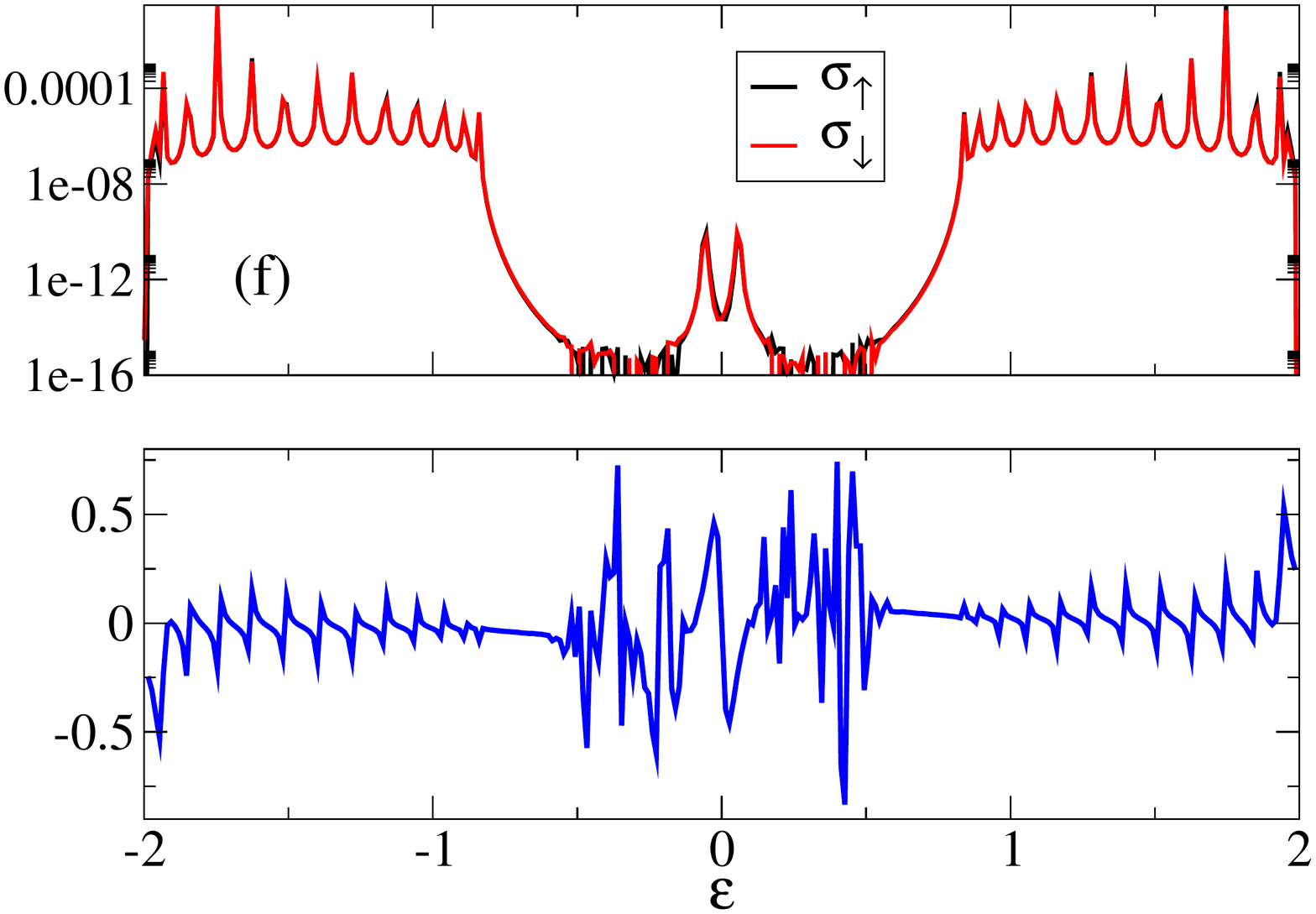}
\caption{\label{size}
Comparison of different numbers of cells and coupling to the leads, for $\gamma=4$.
In the top row the number of cells is $10$ and in the lower row is $20$.
We consider $\lambda=2,\lambda_R=2,B_x=0.1$. We take a) $V=1$, b) $V=0.1$, c) $V=0.05$
d) $V=0.5$, e) $V=0.1$ and f) $V=0.05$.
}
\end{figure*}

Here we disregard the spin content of the electrons and consider that at each location
there are two states, labeled by $m_l=1,2$.
The Hamiltonian in real space is chosen as
\bea
H &=&  \sum_j \sum_{m_l} \left(-\mu+\Delta(m_l) \right) c_{j,m_l}^{\dagger} c_{j,m_l}
\nonumber \\
&-& t_1 \sum_j \sum_{m_l} \left( c_{j,m_l}^{\dagger} c_{j+1,m_l} + c_{j+1,m_l}^{\dagger} c_{j,m_l} \right)
\nonumber \\
&-& t_{\perp} \sum_j \left( c_{j,2}^{\dagger} c_{j,1} + c_{j,1}^{\dagger} c_{j,2} \right)
\nonumber \\
&-& t_2 \sum_j \left( c_{j+1,2}^{\dagger} c_{j,1} + c_{j,1}^{\dagger} c_{j+1,2} \right)
\nonumber \\
&-& t_2 \sum_j \left( e^{-i 2 \pi/\gamma} c_{j-1,2}^{\dagger} c_{j,1} +
e^{i 2 \pi/\gamma} c_{j,1}^{\dagger} c_{j-1,2} \right)
\nonumber \\
\label{Horbital}
\eea
The local hopping between the two orbital states $t_{\perp}$ lifts the degeneracy of the orbital bands,
if large enough. Also, we consider $\Delta(1)=-1, \Delta(2)=1$ and $\mu=0$. The action
of the orbital chiral coupling, $t_2$, is determinant. If the phase of the nearest-neighbor
hopping $t_2$ vanishes, the bands are gapless. In order to look for topological bands we focus on
bands that have gaps. Taking $t_{\perp}=1$ we get a set of bands as shown in Fig. \ref{orbital}.
We consider, as an example, $\gamma=4$ and therefore we have eight bands, since $m_l=1,2$. The bands are separated by
finite gaps. Some of the bands have finite
Berry phases of $\gamma_B=\pi$.

\section{Spin transport}

\subsection{Transfer matrix and scattering states coefficients}

Consider an electron with spin up coming from the left lead, with a given energy $\epsilon$.
We choose
\begin{equation}
\label{leftup} 
\left(\begin{array}{c}
u_{\uparrow}(N_L) \\
u_{\downarrow}(N_L)
\end{array}\right)
= 
\left(\begin{array}{c}
1 \\
0
\end{array}\right)
+r \left(\begin{array}{c}
1 \\
0
\end{array}\right)
+\bar{r} \left(\begin{array}{c}
0 \\
1
\end{array}\right)
\end{equation}
where $r$ is the reflection coefficient and $\bar{r}$ the reflection coefficient with spin flip, and
\begin{equation}
\label{leftup} 
\left(\begin{array}{c}
u_{\uparrow}(N_L-1) \\
u_{\downarrow}(N_L-1)
\end{array}\right)
= 
\left(\begin{array}{c}
1 \\
0
\end{array}\right) e^{-i q_1}
+r \left(\begin{array}{c}
1 \\
0
\end{array}\right) e^{iq_1}
+\bar{r} \left(\begin{array}{c}
0 \\
1
\end{array}\right) e^{iq_1}
\end{equation}
where $q_1=\cos^{-1}\left( -\epsilon/(2 w_2)\right)$ and $w_2$ is the hopping on the left (and right) leads.

Considering an electron with spin down coming from the left lead, with a given energy $\epsilon$, leads to
\begin{equation}
\label{leftup} 
\left(\begin{array}{c}
u_{\uparrow}(N_L) \\
u_{\downarrow}(N_L)
\end{array}\right)
= 
\left(\begin{array}{c}
0 \\
1
\end{array}\right)
+r \left(\begin{array}{c}
0 \\
1
\end{array}\right)
+\bar{r} \left(\begin{array}{c}
1 \\
0
\end{array}\right)
\end{equation}
and
\begin{equation}
\label{leftup} 
\left(\begin{array}{c}
u_{\uparrow}(N_L-1) \\
u_{\downarrow}(N_L-1)
\end{array}\right)
= 
\left(\begin{array}{c}
0 \\
1
\end{array}\right) e^{-i q_1}
+r \left(\begin{array}{c}
0 \\
1
\end{array}\right) e^{iq_1}
+\bar{r} \left(\begin{array}{c}
1 \\
0
\end{array}\right) e^{iq_1}
\end{equation}

On the right lead, we have the transmited wave.
In the case of an electron with spin up coming from the left lead, with a given energy $\epsilon$, we have
\begin{equation}
\label{leftup} 
\left(\begin{array}{c}
u_{\uparrow}(N_R) \\
u_{\downarrow}(N_R)
\end{array}\right)
= 
t \left(\begin{array}{c}
1 \\
0
\end{array}\right)
+\bar{t} \left(\begin{array}{c}
0 \\
1
\end{array}\right)
\end{equation}
where $t$ is the transmission coefficient and $\bar{t}$ the transmission coefficient with spin flip, and
\begin{equation}
\label{leftup} 
\left(\begin{array}{c}
u_{\uparrow}(N_R+1) \\
u_{\downarrow}(N_R+1)
\end{array}\right)
= 
t \left(\begin{array}{c}
1 \\
0
\end{array}\right) e^{i q_1}
+\bar{t} \left(\begin{array}{c}
0 \\
1
\end{array}\right) e^{iq_1}
\end{equation}

Considering an electron with spin down coming from the left lead, with a given energy $\epsilon$, leads to
\begin{equation}
\label{leftup} 
\left(\begin{array}{c}
u_{\uparrow}(N_R) \\
u_{\downarrow}(N_R)
\end{array}\right)
= 
t \left(\begin{array}{c}
0 \\
1
\end{array}\right)
+\bar{t} \left(\begin{array}{c}
1 \\
0
\end{array}\right)
\end{equation}
and
\begin{equation}
\label{leftup} 
\left(\begin{array}{c}
u_{\uparrow}(N_R+1) \\
u_{\downarrow}(N_R+1)
\end{array}\right)
= 
t \left(\begin{array}{c}
0 \\
1
\end{array}\right) e^{i q_1}
+\bar{t} \left(\begin{array}{c}
1 \\
0
\end{array}\right) e^{iq_1}
\end{equation}

The normalization (conservation of probability/charge) implies
\bea
1 &=& |r_{\uparrow}|^2 + |\bar{r}_{\uparrow}|^2 + |t_{\uparrow}|^2+|\bar{t}_{\uparrow}|^2
\nonumber \\
1 &=& |r_{\downarrow}|^2 + |\bar{r}_{\downarrow}|^2 + |t_{\downarrow}|^2+|\bar{t}_{\downarrow}|^2
\eea

Explicitly, the transfer matrix is given by
\begin{equation}
\label{mn2} 
M_n=
\left(\begin{array}{cccc}
C & D \\
I & 0
\end{array}\right)
\end{equation}
with
\begin{equation}
C=
\left(\begin{array}{cc}
\frac{-\left( B_z+\epsilon \right) w - \left( B_x+i B_y \right) \lambda \psi_n}{w^2+\lambda^2} & 
\frac{-\left(B_x-i B_y \right) w -\left(-B_z+\epsilon \right) \lambda \psi_n}{w^2+\lambda^2} \\
\frac{\left( B_z+\epsilon \right) \lambda \psi_n^* - \left( B_x+i B_y \right) w}{w^2+\lambda^2} & 
\frac{\left(B_x-i B_y \right) \lambda \psi_n^* -\left(-B_z+\epsilon \right) w}{w^2+\lambda^2} 
\end{array}\right)
\end{equation}
and
\begin{equation}
D=
\left(\begin{array}{cc}
\frac{-w^2 +\lambda^2 \psi_n \psi_{n-1}^*}{w^2+\lambda^2} &
\frac{-w \lambda \psi_{n-1} -w \lambda \psi_n}{w^2+\lambda^2} \\
\frac{w \lambda \psi_{n}^* +w \lambda \psi_{n-1}^*}{w^2+\lambda^2} &
\frac{-w^2 +\lambda^2 \psi_n^* \psi_{n-1}}{w^2+\lambda^2} \\
\end{array}\right)
\end{equation}
$I$ is the identity and $0$ the null matrix.
The coefficients, for a scattering state in a spin up incident electron,  may be obtained solving the equations
\cite{houston}
\begin{equation}
\label{leftup} 
A_{\uparrow}
\left(\begin{array}{c}
r \\
t \\
\bar{r} \\
\bar{t}
\end{array}\right)
= b_{\uparrow}
\end{equation}
where
\begin{equation}
\label{leftup} 
b_{\uparrow}=-
\left(\begin{array}{c}
M_{11}^T+M_{13}^T e^{-i q_1} \\
M_{21}^T+M_{23}^T e^{-i q_1} \\
M_{31}^T+M_{33}^T e^{-i q_1} \\
M_{41}^T+M_{43}^T e^{-i q_1}
\end{array}\right)
\end{equation}
and
\begin{equation}
\label{mn2} 
A_{\uparrow}=
\left(\begin{array}{cccc}
M_{11}^T+M_{13}^T e^{i q_1} & -e^{i q_1} & M_{12}^T+M_{14}^T e^{i q_1} & 0 \\
M_{21}^T+M_{23}^T e^{i q_1} & 0 & M_{22}^T+M_{24}^T e^{i q_1} & -e^{i q_1} \\
M_{31}^T+M_{33}^T e^{i q_1} & -1 & M_{32}^T+M_{34}^T e^{i q_1} & 0 \\
M_{41}^T+M_{43}^T e^{i q_1} & 0 & M_{42}^T+M_{44}^T e^{i q_1} & -1 \\
\end{array}\right)
\end{equation}
and similarly for a spin down incident electron.


\subsection{Influence of chain size and coupling to leads}

As mentioned before, the detection of the edge localized states is enhanced
considering relatively small systems, so that the exponential decay of the 
wave functions is not significative. Also, the smallness of the coupling to the leads
is required otherwise the edge states do not give a significant contribution to the
differential conductance.

Some examples of the balance of these dependencies is considered in Fig. \ref{size} considering
as an example two chains with $10$ and $20$ unit cells and different couplings $V$.
As the results show the spin polarization is more sensitive to the edge states,
since it is given by a ratio of coefficients as shown in Eq. (\ref{definition}).
Intermediate values of the coupling, $V=0.1,0.5$, for the system sizes considered
give reasonable results.


\begin{thebibliography}{}


\bibitem{bauer}
T. Yu, Z. Luo and G. E. W. Bauer, arXiv:2206.05535.

\bibitem{xie}
Z. Xie, T.Z. Markus, S. R. Cohen, Z. Vager, R. Gutierrez, and R. Naaman, Nano Lett. {\bf 11}, 4652 (2011).

\bibitem{guo1}
A.- M. Guo, and Q.- F. Sun, Phys. Rev. Lett. {\bf 108}, 218102 (2012).

\bibitem{gutierrez1}
R. Gutierrez, E. D\'{\i}az, R. Naaman, and G. Cuniberti, Phys. Rev. B {\bf 85}, 081404(R) (2012).

\bibitem{eremko}
A. A. Eremko and V. M. Loktev, Phys. Rev. B {\bf 88}, 165409 (2013).

\bibitem{gutierrez2}
R. Gutierrez, E. D\'{\i}az, C. Gaul, T. Brumme, F. Dom\'{\i}nguez-Adame, and G. Cuniberti,
J. Phys. Chem. C {\bf 117}, 22276 (2013).


\bibitem{gutierrez3}
R. Gutierrez, and G. Cuniberti, Acta Phys. Polon. A {\bf 127}, 185 (2015).

\bibitem{naaman}
R. Naaman and D. H. Waldeck, Annu. Rev. Phys. Chem. {\bf 66}, 263 (2015).

\bibitem{varela}
S. Varela, V. Mujica, and E. Medina, Phys. Rev. B {\bf 93}, 155436 (2016).

\bibitem{caetano}
R. A. Caetano, Sci. Reports {\bf 6}, 23452 (2016).

\bibitem{matityahu}
S. Matityahu, Y. Utsumi, A. Aharony, O. E.- Wohlman, and C. A. Balseiro, Phys. Rev. B {\bf 93}, 075407 (2016).

\bibitem{utsumi}
Yasuhiro Utsumi, Ora Entin-Wohlman, and Amnon Aharony
Phys. Rev. B {\bf 102}, 035445 (2020).

\bibitem{inui}
A. Inui, R. Aoki, Y. Nishiue, K. Shiota, Y. Kousaka, H. Shishido, D. Hirobe, M. Suda, J.-ichiro Ohe, 
J-ichiro. Kishine,
H. M. Yamamoto, and Y. Togawa, Phys. Rev. Lett. {\bf 124}, 166602 (2020)

\bibitem{evers}
Ferdinand Evers, Amnon Aharony, Nir Bar-Gill, Ora Entin-Wohlman, Per Hedeg\"ard, 
Oded Hod, Pavel Jelinek, Grzegorz Kamieniarz, Mikhail Lemeshko, Karen Michaeli, 
Vladimiro Mujica, Ron Naaman, Yossi Paltiel, Sivan Refaely-Abramson, Oren Tal, 
Jos Thijssen, Michael Thoss, Jan M. van Ruitenbeek, Latha Venkataraman, 
David H. Waldeck, Binghai Yan, and Leeor Kronik,
Adv. Mater. {\bf 34}, 2106629 (2022), arXiv:2108.09998.

\bibitem{liu}
Y. Liu, J. Xiao, J. Koo, and B. Yan, Nature Materials {\bf 20}, 638 (2021), arXiv:2008.08881 (2020).

\bibitem{fransson}
J. Fransson, Nano Lett. {\bf 21}, 3026 (2021), arXiv:2103.00840.

\bibitem{lorman1}
V.L. Lorman and B. Mettout, Phys. Rev. Lett. {\bf 82}, 940 (1999).

\bibitem{selinger}
A. J\'akli, O. D. Laventrovich and J. V. Selinger, Rev. Mod. Phys {\bf 90}, 045004 (2018).

\bibitem{lorman2}
V.L. Lorman and B. Mettout, Phys. Rev. E {\bf 69}, 061710 (2004).

\bibitem{mettout1}
B. Mettout, Phys. Rev. E {\bf 75}, 011706 (2007).

\bibitem{SSH}
W. P. Su, J. R. Schrieffer, and A. J. Heeger, Phys. Rev. Lett.
\textbf{42}, 1698 (1979).

\bibitem{Verresen}
R. Verresen, N. G. Jones and F. Pollmann,
Phys. Rev. Lett. {\bf 120}, 057001 (2018).

\bibitem{prr3}
O. Balabanov, D. Erkensten and H. Johannesson,
Phys. Rev. Research {\bf 3}, 043048 (2021).

\bibitem{ssh1}
Linhu Li, Zhihao Xu, and Shu Chen,
Phys. Rev. B {\bf 89}, 085111 (2014).

\bibitem{ssh2}
Zhi-Hai Liu, O. Entin-Wohlman, A. Aharony, J. Q. You, and H. Q. Xu,
Phys. Rev. B {\bf 104}, 085302 (2021).

\bibitem{guo2}
A.- M. Guo, and Q.- F. Sun, Phys. Rev. B {\bf 95}, 155411 (2017).

\bibitem{schuster}
Thomas Schuster, Felix Flicker, Ming Li, Svetlana Kotochigova,	Joel E. Moore,	Jun Ye,	and Norman Y. Yao,
Phys. Rev. A {\bf 103}, 063322 (2021).

\bibitem{zak} 
J. Zak, Phys. Rev. Lett. {\bf 62}, 2747 (1989).

\bibitem{pareek}
T. P. Pareek and P. Bruno, Phys. Rev. B {\bf 65}, 241305(R) (2002).

\bibitem{hasan}
M. Z. Hasan and C. L. Kane, Rev. Mod. Phys. {\bf 82}, 3045 (2010).

\bibitem{zhang}
X.- L- Qi and S. C. Zhang, Rev. Mod. Phys. {\bf 83}, 1057  (2011).

\bibitem{kane}
C. L. Kane and E. J. Mele, Phys. Rev. Lett. {\bf 95}, 146802 (2005);
Phys. Rev. Lett. {\bf 95}, 226801 (2005).

\bibitem{gentile}
P. Gentile, M. Cuoco and C. Ortix, Phys. Rev. Lett. {\bf 115}, 256801 (2015).

\bibitem{malard}
Mariana Malard, David Brandao, Paulo Eduardo de Brito, and Henrik Johannesson,
Phys. Rev. Research {\bf 2}, 033246 (2020).

\bibitem{schnyder}
A.P. Schnyder, S. Ryu, A. Furusaki and A.W.W. Ludwig,
Phys. Rev. B {\bf 78}, 195125 (2008).



\bibitem{benalcazar} 
W. A. Benalcazar, B. A. Bernevig and T. L. Hughes, Phys. Rev. B {\bf 96}, 245115 (2017).

\bibitem{niu}
D. Xiao, M.-C. Chang and Q. Niu, Rev. Mod. Phys. {\bf 82}, 1959 (2010).

\bibitem{resta}
R. Resta, J. Phys. Condens. Matter {\bf 22}, 123201 (2010).

\bibitem{fukui}
T. Fukui, Y. Hatsugai and H. Suzuki, J. Phys. Soc. Japan {\bf 74}, 1674 (2005).

\bibitem{vanderbilt}
R. D. King-Smith and D. Vanderbilt, Phys. Rev. B {\bf 47}, 1651 (1993).

\bibitem{resta2}
R. Resta, Rev. Mod. Phys. {\bf 66}, 899 (1994).

\bibitem{bahari}
M. Bahari and M. V. Hosseini, Phys. Rev. B {\bf 94}, 125119 (2016).


\bibitem{dong}
B. Dong and X.L. Lei,
Annals of Phys. {\bf 396}, 245 (2018). 

\bibitem{balabanov}
Oleksandr Balabanov, and Henrik Johannesson,
J. Phys.: Condens. Matter {\bf 32}, 015503 (2020).

\bibitem{Landauer}
R. Landauer, Philos. Mag. {\bf 21}, 863 (1970);
D. C. Langreth and E. Abrahams, Phys. Rev. B {\bf 24}, 2978 (1981);
M. B\"uttiker, Y. Imry, R. Landauer and S. Pinhas, Phys. Rev. B {\bf 31},
6207 (1985);
M. B\"uttiker and T.M. Klapwijk, Phys. Rev. B {\bf 33}, 5114 (1986).

\bibitem{houston}
J.-X. Zhu and C.S. Ting, Phys. Rev. B {\bf 61}, 1456 (2000).

\bibitem{creutz}
M. Creutz, 
Phys. Rev. Lett. {\bf 83}, 2636 (1999).

\bibitem{platero}
Juan Zurita, Charles Creffield, and Gloria Platero,
Quantum {\bf 5}, 591 (2021).

\end{thebibliography}
\end{document}